# Expecting to be HIP: Hawkes Intensity Processes for Social Media Popularity

Marian-Andrei Rizoiu†♯, Lexing Xie†♯, Scott Sanner‡
Manuel Cebrian♯, Honglin Yu†♯, Pascal Van Henteryck♭
†Australian National University, ♯Data61 CSIRO, ‡University of Toronto, ♭University of Michigan

## ABSTRACT

Modeling and predicting the popularity of online content is a significant problem for the practice of information dissemination, advertising, and consumption. Recent work analyzing massive datasets advances our understanding of popularity, but one major gap remains: To precisely quantify the relationship between the popularity of an online item and the external promotions it receives. This work supplies the missing link between exogenous inputs from public social media platforms, such as Twitter, and endogenous responses within the content platform, such as YouTube. We develop a novel mathematical model, the Hawkes intensity process, which can explain the complex popularity history of each video according to its type of content, network of diffusion, and sensitivity to promotion. Our model supplies a prototypical description of videos, called an *endo-exo map*. This map explains popularity as the result of an extrinsic factor – the amount of promotions from the outside world that the video receives, acting upon two intrinsic factors – sensitivity to promotion, and inherent virality. We use this model to forecast future popularity given promotions on a large 5-months feed of the most-tweeted videos, and found it to lower the average error by 28.6% from approaches based on popularity history. Finally, we can identify videos that have a high potential to become viral, as well as those for which promotions will have hardly any effect.

## 1. INTRODUCTION

The popularity of an online cultural item is described by the amount of attention it receives, and the popularity dynamics refers to its evolution over time. Popularity is a critical measure of information dissemination for content producers, and a way to manage information overload for content consumers. Understanding and predicting popularity have been active topics in both research and practice, but many fundamental questions remain open, such as: What describes the most viral items? What do the popularity dynamics of news, music, films look like, and what are their differences and similarities? Can we promote an item to increase its popularity, and how much promotion is needed?

Building upon recent research progress in understanding popularity, we identify three important questions that are still open. The first one concerns modeling popularity. One set of approaches describe popularity dynamics as stylistic prototypes, such as being power-law shapes from either an exogenous shock or endogenous relaxation [13], a combination of power-law and exponential decay [24], multiple power-law decays with periodicity [27] or a collection of recurrence peaks [10]. However, one question remains: **How would popularity evolve under continuous external influence?** Especially, how one can explain complex rise and fall patterns that do not follow the prescribed prototypes. The second questions concerns virality. Content and initial diffusion have both been identified as key factors that influence popularity. Here content factors include positive sentiment [2], emotional arousal [5], publishing venue [3], visibility [6]; and factors of diffusion history include [9] network structure, information about the original poster and re-sharers, the timing of the early posts. However, describing viral content in the light of external promotions is still an open problem, and in particular: **Can something go viral if promoted?** The third questions involves predicting future popularity. It is known that the approaches that use the popularity history [30, 34] produce competitive estimates about future popularity over time. Also, timing features have been shown to be more predictable than content, structure, and user features [9], and prediction without initial history is generally shown as a hard problem [26]. However, these recent insights do not answer: **How to forecast future popularity given planned promotions?**

In this work, we answer all three questions above, using a large dataset that connects popularity in one social media platform – 81.9 million YouTube videos – to discussions about each of these digital items in an external platform – in 1.06 billion tweets over a six-month period.

To describe complex popularity dynamics under continuous external influence, we propose a new mathematical model that reveals an analytical relationship between endogenous and exogenous demand factors, called the Hawkes Intensity Process (HIP). HIP extends the well-known Hawkes point-process [19], by taking the expectation over stochastic event histories so as to describe expected event volumes, rather than a set of event times. Figure 1 illustrates the HIP model. On the top left is the volume of exogenous promotions over time, which drives the endogenous response determined by the HIP (middle); the output on the right is the popularity series. The popularity series modeled through the Hawkes intensity process matches closely with the observed view count series, even for videos with complex popularity lifecycles (Section 2).

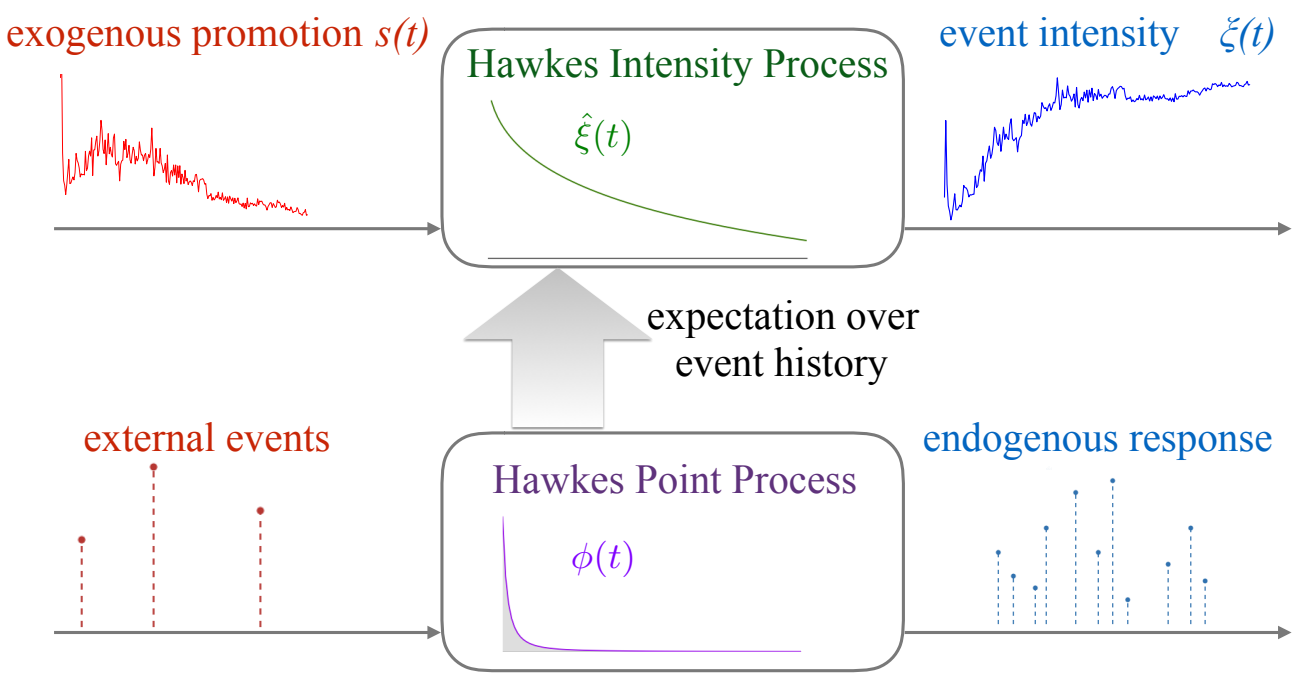

Figure 1: Linking endogenous and exogenous factors of popularity using the Hawkes Intensity Process. *Top row*: The input are volumes of exogenous promotion or discussions $s(t)$, that engender endogenous reactions from the online social networks described by the impulse response function $\hat{\xi}(t)$ (middle box, defined in Sec 2.5), to generate the total popularity series $\xi(t)$. *Bottom row*: The endogenous reactions are self-exciting point processes, widely used in recent literature [4, 23, 28, 31, 33, 39]. Here each event triggers subsequent events with memory kernels $\phi(t)$. Such point process models can incorporate individual external stimulus (show on the left) which in turn lead to a larger number of events in response (shown on the right). *Middle arrow*: The proposed HIP model is a result of taking the expectation over all stochastic event history of the Hawkes process in the bottom.

To answer the second question, on whether or not an item will go viral if promoted, we derive two new metrics based on HIP – the endogenous response and exogenous sensitivity. These two metrics naturally lend to a novel two-dimensional visualization tool, dubbed the *endo-exo map* (Section 4). On this map, one can identify online videos that have high potential but are not yet popular. In other words, video with high sensitivity to external promotions and high endogenous response are expected to go viral if promoted. On the other hand, one can also identify videos for which promotion is unlikely to have an effect, such as those scoring very low in either the endo- or exo- dimension.

Finally, the HIP model can be used to help forecast future popularity given (known or planned) promotions. HIP model parameters are estimated on the first 90 days of each video's history, and forecasts are made for the next 30 days. We evaluate forecasting on a collection of 13K+ most actively discussed YouTube videos over a six-month period, and found that estimates made with the HIP lower the average percentile error by 28.6% from state-of-the-art methods based on popularity history (Section 5).

The main contributions of this work include:

- The HIP model, a volume based version of the Hawkes point process. Its essential novelty is to regard popularity as externally-driven, with exogenous events activating endogenous responses inside the social environment which may, or may not, amplify the exogenous signal.
- The exogenous sensitivity and the endogenous response, two new metrics to quantify two distinct aspects of a video's inherent tendency to be popular. They are combined in the endo-exo map, a tool used to comparatively explain popularity and identify potentially viral videos.
- A method to forecast popularity gain after promotion. Evaluated on a large set of YouTube videos, it significantly outperforms approaches using popularity history.
- A new dataset of tweeted videos that links online videos to their external discussions, available at https://github.com/andrei-rizoiu/hip-popularity.

# 2. THE MODEL

We introduce a model for the evolution of online attention under external influence. We start by discussing the problem setting of aggregated attention under external promotion (in Sec. 2.1), the key concepts of the Hawkes process and its use to link the ongoing effect of external stimuli to the word-of-mouth spread of attention (Sec. 2.2). Next, we propose HIP, a model to explain the observed popularity history from daily volumes when the underlying viewing events are unobserved (Sec. 2.3). Lastly, we introduce two key metrics derived from the HIP model, the endogenous response and exogenous sensitivity, to quantify the viral potential of a video (Sec. 2.5).

## 2.1 Problem setting: views under promotion

This paper aims to model the popularity of videos under external promotion. Here popularity is measured in the number of total views after the video being online for a certain number of days (e.g. up to 120 days). External promotion is harder to measure, since by definition, it needs to capture data from other platforms. In this paper, we have two different views of promotion, due to the data collection setting described in Sec 3. The first is *shares*, tracked by YouTube via the *share* button under each video that allows a user to share a link of the video on a selection of popular social network sites – 13 at the time of this writing. The second view is *tweets*, tracked with twitter streaming API with keyword filters that retrieve tweets that link to a video. Neither source is complete – with the distributed nature of the Internet, one can see that a complete capture of all discussions is practically impossible. The *shares* captures external promotions from a diverse set of sources, but is far from complete in any one source. The *tweets* captures an almost-complete feed of video promotions in one platform. In the rest of this paper, both of these sources are collectively referred to as external *promotions* about a video. In our evaluations, the results obtained using each source are presented separately.

## 2.2 Hawkes process for social events

We model online attention as an exogenously-driven self-exciting process – each viewing *event* is triggered either by a previous event or as a result of external influence. We assume that viewing events of a YouTube video follow a Hawkes point process [19], a type of non-homogeneous point process in which the arrival of an event increases the likelihood of future events. Although variants of point processes have recently been used to model events in social media, all existing work focus on learning point process model from one information source, such retweeting [39, 23], arrival of citations [33], or endogenous response after an initial external shock [13]. To the best of our knowledge, this is the first work that models the continuous interaction of two sources – exogenous stimuli and endogenous response.

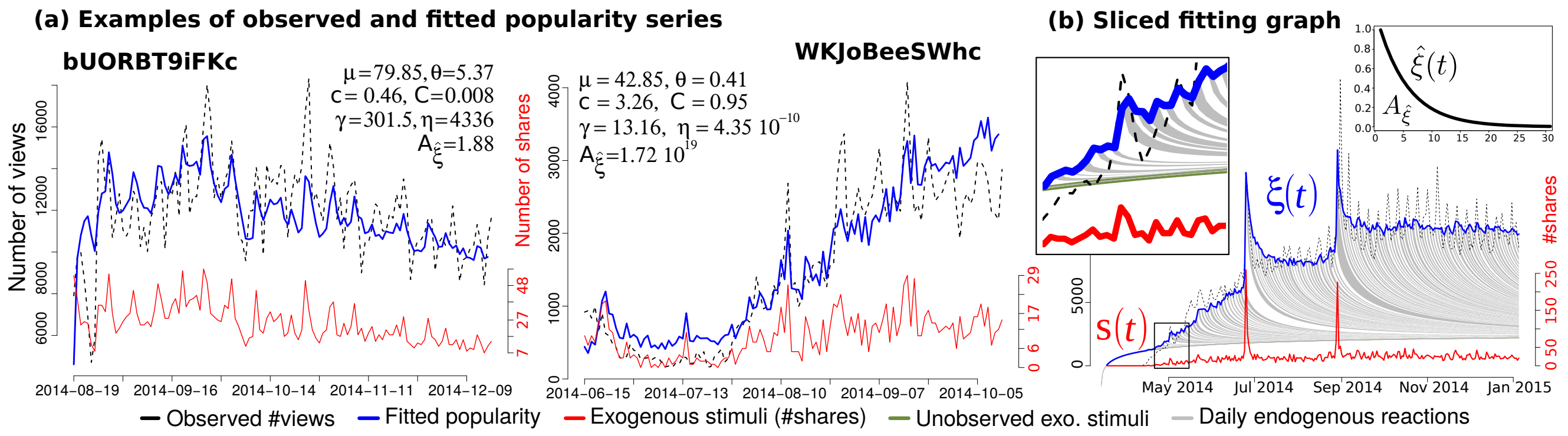

Figure 2: Explaining popularity dynamics using the Hawkes intensity model. **(a)** Number of shares (red), observed popularity history (black dashed) and popularity as explained by the HIP (blue) on two examples videos: a `music` video *bUORBT9iFKc* and a `News & Politics` video *WKJoBeeSWhc*. The multi-phased popularity history cannot be explained by current models such as [13], while the HIP tracks the complex dynamics well. **(b)** A *sliced* fitting graph of a music video (Youtube ID *0bR4L0Y94AQ*) – using the impulse response $\hat{\xi}(t)$ and exogenous stimuli $s(t)$ to explain observed popularity. Each alternating gray and white area under the fitted (blue) curve is a *slice* of endogenous reaction generated by the external influence in a given day. The left inset zooms-in one of the early months in the video's evolution, in May 2014. The total event intensity (blue solid line) is a sum of temporally shifted and scaled versions of $\hat{\xi}(t)$, which tracks the long-term trends in observed popularity well (dashed line). The period around the first larger exogenous peak is shown magnified so that its corresponding endogenous response is clearly visible. (right inset) Example of the impulse response $\hat{\xi}(t)$ to one unit of external excitation. The area under this function, $A_{\hat{\xi}}$, quantifies the endogenous reaction of a video – it is the total number of views after each unit of exogenous excitation.

In particular, the arrival rate of viewing events $\lambda(t)$, a measure of how likely a viewing event will occur in a infinitesimal interval around time $t$, is determined by two additive components in Eq (1). The first component is proportional to a measure of external influence $s(t)$ scaled by a constant $\mu$. Here $s(t)$ represents the volume of external discussion (or promotion) over time. The second component represents the rate of views triggered by a previous event $i$, which occurred at time $t_i$ with magnitude $m_i > 0$, according to a time-decaying triggering kernel $\phi_{m_i}(t - t_i)$. Furthermore, each event $t_i < t$ adds to $\lambda(t)$ independently. The following equations describe the event rate of such a *marked* Hawkes process:

$$\lambda(t) = \mu s(t) + \sum_{t_i<t} \phi_{m_i}(t - t_i) \quad (1)$$

$$\phi_m(\tau) = \kappa\, m^\beta\, (\tau + c)^{-(1+\theta)},\ \tau \in \mathcal{R}^+ \quad (2)$$

Eq. 2 describes the triggering kernel $\phi(\tau)$. In this work it is designed to capture several key quantities influencing popularity. Parameter $\kappa$ is a scaling factor for video *quality*. $m$ describes the relative *influence* of the user who generated the event, i.e., $m_i$ in Eq. 1 when multiple events are concerned. The user influence exponent $\beta$, newly introduced in this work, accounts for the nonlinearity between observed metrics of influence (such as the number of followers) and popularity. This particular form allows both flexibility in modeling how much effect some observed metric of *influence* (e.g. number of followers) has on views (e.g. $\beta = 0$ would be no effect), and at the same time computing expectations over stochastic event history analytically, as will be shown in the next subsection. Time interval $\tau = t - t_i$ is the elapsed time since the parent event at $t_i$; $c > 0$ is a cutoff term to keep $\phi_m(\tau)$ bounded when $\tau$ is small; $1 + \theta$ (for $\theta > 0$) is the power-law exponent for social memory – the larger $\theta$ is, the sooner the reaction to an event will stop. We use a power-law kernel for $\phi_m(\tau)$, as recent work [28] observed it to have better performance on social media data than popular variants like the exponential kernel.

This model is an instance of a marked Hawkes process [19]. An illustration of the Hawkes process with external excitation is in the bottom row of Figure 1. A set of input events of different magnitudes trigger new events through the kernel $\phi(t)$, which then trigger offspring events themselves, resulting in the observed event sequence.

## 2.3 From Hawkes to HIP

The Hawkes point-process faces a few modeling challenges in large-scale applications. In terms of data source, what we often observe is the volume of total attention in a given interval (e.g. daily views on YouTube), rather than the times and properties of individual actions, due to constraints in user privacy and data volume. In terms of computation, full estimation of the Hawkes process is quadratic in the number of events. Therefore, the full estimation quickly becomes expensive when the number of events is in the hundred thousands or millions – this is where the most popular videos are (see Sec 3.2). It is very desirable if one could estimate video popularity with daily data, which is typically a few dozens to a few hundred data points.

To this end, we introduce the Hawkes intensity $\xi(t)$, the expectation of the event rate $\lambda(t)$ over the event history $\mathcal{H}_t$, consisting of the set of (random) event times and magnitudes up to time $t$.

THEOREM 2.1. HAWKES INTENSITY PROCESS (HIP) *Given a marked Hawkes process described in Equations (1) and (2). Its event history*

$$\mathcal{H}_t = \{(t_1, m_1), \ldots, (t_n, m_n)\}_{t_n<t}$$

*contains all event times and marks before time t, where each mark m is drawn iid from a power-law distribution* $p(m) = (\alpha-1)m^{-\alpha}$. *We define event intensity as the expectation of the event rate over the event history* $\xi(t) = \mathbb{E}_{\mathcal{H}_t}[\lambda(t)]$, *then*

*$\xi(t)$ follows the following self-consistent integral equation:*

$$\xi(t) = \mu s(t) + C\int_0^t \xi(t-\tau)(\tau + c)^{-(1+\theta)}d\tau \ . \qquad (3)$$

*Here constant $C = \frac{\kappa(\alpha-1)}{\alpha-\beta-1}$, and $\kappa$ and $\beta$ are as in Eq (2).*

Intuitively, this expression of event intensity $\xi(t)$ at time $t$ is determined by the external stimulus $s(t)$, and a convolution of its own history with a power-law memory kernel $(\tau+c)^{-(1+\theta)}$. Theorem 2.1 can be intuitively understood by breaking down the expectation into several parts. Note $\mu s(t)$ is non-random and does not change after expectation. We compute analytically the expectation over stochastic history, with a random number of events at random times, by decomposing $E_{\mathcal{H}_t}$ into expectations over binary variables $dNt$, which indicates whether or not there is an event in a small interval around time $t$. This trick discretizes time, and converts the sum over past events in Eq (1) into an integration seen in Eq (3). Note that the expectation of the user influence warping term $m^\beta$ over the power-law distribution of the mark $m$ has an analytical form, leading to the constant $C$. Due to space limitations, we include the full proof in the online appendix [1].

Here $(\mu, \theta, C, c)$ are video-dependent parameters estimated from the popularity history of each video. Note that $\alpha > 0$ is the power-law exponent of user influence distribution, estimated as $\alpha = 2.016$ from a large Twitter sample using standard fitting procedures [11]. The two power law exponents $\alpha$ and $\theta$ in HIP are distinct in meaning and function, $\theta$ defines memory decay over time, while $\alpha$ is determined by the user distribution at large.

Compared to existing models of data volume, HIP captures the ongoing interactions of exogenous and endogenous effects. Hence it is able to explain complex popularity series with multiple rises and falls (as shown in Figure 2). Helmstetter and Sornette [20] fit the observed event rate after an initial shock, and Crane and Sornette [13] produce a curve fit on the long-term approximation of the endogenous decay with no exogenous input. SpikeM [27] models volumes of events both prior and after a single considered shock, without accounting for external influences. The work most related to ours on computing expectations over stochastic event histories is th work of Farajtabar *et al.* [16], who modeled co-excitation on Twitter and computed the equivalent of $\xi(t)$ on multivariate Hawkes process with exponential kernels, which admits a closed-form solution. In contrast, our work uses a univariate Hawkes process focused on modeling the impact of Twitter on individual Youtube videos and a power law kernel. De *et al.* [14] further develop the work in [16] by combining a Markov process with a multivariate Hawkes process for modeling opinion dynamics.

## 2.4 Estimating HIP from data

We discuss key steps for estimating HIP from observed series of views and external promotions over time.

**Discretizing over time**. We observe that behavioral statistics are aggregated over fixed and discrete intervals – for YouTube, the public API provides the daily history of the number of views $\bar{\xi}[t]$ and number of shares $\bar{s}[t]$ for $t = 1, \ldots, T$. Expressing HIP (Eq (3)) over discrete time gives:

$$\xi[t] = \mu s[t] + C\sum_{\tau=1}^{t} \xi[t-\tau](\tau + c)^{-(1+\theta)} \ . \qquad (4)$$

Here we use square brackets to denote discrete time, e.g. $\xi[t]$, and round brackets to denote continuous time, e.g. $\xi(t)$.

**Accounting for unobserved external influence.** In addition to the observed external promotions $\bar{s}[t]$ in tweets or shares, we model the unobserved external excitation as an initial shock (at $t = 0$) and a constant background excitation (for $t > 0$).

$$s[t] = \frac{\gamma}{\mu}\mathbb{1}[t=0] + \frac{\eta}{\mu}\mathbb{1}[t>0] + \bar{s}[t] \ , \qquad (5)$$

where $\mathbb{1}(arg)$ is the standard impulse function – taking the value 1 when $arg$ is true and 0 otherwise. In the absence of a parametric model of generic external influence, the initial impulse and the constant component require the least amount of assumptions about how unobserved influence evolves. Here $\gamma$ and $\eta$ are additional parameters estimated from data. In our experiments, adding estimates for such unobserved influence components improves the fitting for a large number of videos.

**The loss function** For each video, we find an optimal set of model parameters $(\mu, \theta, C, c)$ and of unobserved external influence ($\gamma$ and $\eta$). This is done by minimizing the square error between the observed viewcount series $\bar{\xi}[t]$ and the model $\xi[t]$, $t = 1 : T$. The corresponding optimization problem is as follows:

$$\min_{\mu,\theta,C,c,\gamma,\eta} J = \frac{1}{2}\sum_{t=0}^{T}\left(\xi[t] - \bar{\xi}[t]\right)^2 \qquad (6)$$

We use `L-BFGS` [25] with analytical gradients and random restarts to minimize this non-linear loss function. Gradient computation is detailed in the appendix [1].

Three example fits are shown in Figure 2. Visibly, the event intensity model in Equation 3 links the exogenous and the endogenous effects of the social system, resulting in a tight fit between the model and the observed popularity history. For the Brazilian music video *bUORBT9iFKc* the memory kernel decays fast ($\theta = 5.37$), and the resulting intensity series tracks the temporal dynamics of the stimuli closely. For news video *WKJoBeeSWhc*, the memory kernel decays slowly ($\theta = 0.41$), hence the delayed accumulation of exogenous promotion via the memory kernel results in an overall rising trend. We can see that only by capturing the non-obvious joint effects from within and outside a social network can a model produce both fine-grained short-term dynamics and accurate long-term trends.

## 2.5 Properties of the HIP

In this section, we examine the key property of HIP of being a linear time-invariant system, which leads to two important metrics for measuring two distinct aspects of a video's viral potential – the exogenous sensitivity and the endogenous response.

**Exogenous sensitivity** $\mu$. As shown in Eq 3, the total attention that a video receives consists of two parts: the input from the exogenous stimuli, and the endogenous response corresponding to non-linear effects accumulated through the integral equation. The scaling parameter $\mu$ quantifies a video's sensitivity to external stimuli $s(t)$. When $\mu \to 0$, external promotion would have no effect; when $\mu$ is large, each unit of external promotion leads to a large number of new views.

**HIP as an LTI system**. We observe an important property of the HIP model.

Corollary 2.2. *The HIP model,* as defined in Eq (4) and (3)*, is a linear time-invariant (LTI) system for* $t > 0$.

Being an LTI system [29] is to say that if $\xi[t]$ is the event intensity function for input $s[t]$, then (for the same video) the event intensity function for a shifted and scaled version of the input $as[t - t_0]$ is $a\xi[t - t_0]$ for $a > 0, t_0 \geq 0$, i.e., scaled and shifted by the same amount.

It is easy to see linearity holds by multiplying both sides of Eq 3 by the same constant. For time invariance, change of variable and then using the fact that $\xi[t] = 0$ when $t < 0$. A full proof is in the appendix [1].

**Impulse response function** $\hat{\xi}[t]$. One important descriptor of an LTI system is the impulse response function, the response to the unit impulse function $\mathbb{1}[t]$, which takes the value 1 when $t = 0$, and 0 otherwise. We define $\hat{\xi}[t]$ as the impulse response of the HIP model. It follows from Eq. (4) that $\hat{\xi}[t]$ is the solution to the following self-consistent equation:

$$\hat{\xi}[t] = \mathbb{1}[t] + C \sum_{t=0}^{T} \hat{\xi}[t - \tau](\tau + c)^{-(1+\theta)} \quad , \tag{7}$$

For each video, $\hat{\xi}[t]$ completely characterizes the endogenous response of the HIP model:

Lemma 2.3. Sliced Responses *The intensity function* $\xi[t]$ *of HIP can be written as the sum of impulse responses, scaled and shifted by the corresponding external input.*

$$\xi[t] = \sum_{\tau=0}^{T} s[\tau]\hat{\xi}[t - \tau] \tag{8}$$

To see that this is true, first notice that external input $s[t]$ can be expressed as a sum of shifted and scaled impulses.

$$s[t] = \sum_{\tau=0}^{T} s[\tau]\mathbb{1}[t - \tau] \tag{9}$$

Combining Eq (7) and (9) will lead to Eq (8). In other words, the total popularity at time $T$ can be obtained as the sum of the unfolding through the endogenous reaction, of the external stimuli having occurred at times $1, 2, \ldots, T-1$. Fig 2(b) illustrates this property using a *sliced* and *stacked* popularity graph. The alternating white and gray *slices* are scaled (and shifted) versions of the impulse response represented in the right inset. For each discrete time point $t'$ corresponds a slice, scaled by the external stimuli $s(t')$, which adds to the slices constructed at previous times $t < t'$. Adding all these slices together recovers the overall intensity $\xi(t)$ as in Eq 3 (blue line), which tracks closely the long-term dynamics of the observed popularity (dashed line). The LTI property and its related quantities provides the mathematical ground to define our second important measure.

**Endogenous response** $A_{\hat{\xi}}$. We define the total *endogenous response* generated from a single unit of exogenous excitation, computed as $A_{\hat{\xi}} = \sum_{t=0}^{\infty} \hat{\xi}[t]$. In this work, we compute $A_{\hat{\xi}}$ by taking the sum over 10,000 time steps. $A_{\hat{\xi}}$ is finite when the underlying HIP is so-called *sub-critical*. Other HIP-derived quantities, such as scaling parameter $C$ or memory exponent $\theta$ could potentially serve to describe video virality. We find, however, that despite being related, the non-linear interactions among HIP parameters render them inaccurate in explaining popularity compared to $A_{\hat{\xi}}$. Detailed discussions on the convergence criteria for $A_{\hat{\xi}}$, and visualizations of other parameters are in the appendix [1]. Together with exogenous sensitivity $\mu$, this is the second key quantity for measuring video virality. They will be used to compare individual and collections of videos in Sec. 4.

# 3. THE TWEETED VIDEOS DATASET

A key component in linking the exogenous influence and the endogenous response is to obtain data for the exogenous component, preferably both inside and outside the studied social network. We describe a new dataset across Twitter and Youtube networks, linked via the unique video ids, in which the volumes of tweets and Youtube shares serve as exogenous signals. We then introduce the popularity scale, a mapping between the number of views (or shares, or tweets) and the percentile ranking of a video, which will be used for visualizing popularity and for evaluating popularity forecast.

## 3.1 Dataset construction

We collect a dataset of *tweeted videos* by streaming tweets (via Twitter API) published between 2014-05-29 and 2014-12-26 which mentions YouTube videos. This yields a large and diverse set of over 81.9 million videos mentioned in 1.06 billion tweets. We obtain from YouTube their video metadata, including upload date, author and video category, as well as the time series consisting of the daily number of views and shares. The video categories are a one-level YouTube classification of videos, example of such categories being `Music`, `Gaming` or `Film & Animation`. Along with the daily number of tweets, we have three attention-related time series for each video: ($views[t]$, $shares[t]$ and $tweets[t]$), where $t$ indexes time with the unit of a day.

In order to study videos with non-trivial popularity and promotion activities, we construct a subset, denoted as the Active dataset, by restricting to videos that are still online and that have their popularity and sharing series at least 120 days long, since the upload and until the crawling date. Furthermore, we restrict the set of videos to those that received at least 100 tweets and 100 shares by the 120th day, in order to obtain videos twitted and shared enough to estimate the external influence on popularity. We also remove 6 rare categories containing less than 1% videos (and their corresponding videos). The Active dataset contains 13,738 videos across 14 categories and it is used in both explaining and forecasting popularity in Sec. 5. Reasons for the drastic dataset reduction from 81M to Active include: videos uploaded earlier than 2014-05-29 (and hence without a complete tweet history), videos that are no longer online, those do not make viewcount history public, and the long-tailed distribution of tweets and shares – more than half of the videos are tweeted only once. Note that when they exist, the popularity and the sharing series do not contain missing data. A profile of the tweeted videos dataset and more details about its construction are given in the appendix [1]. We use the first 90 days of each videos' viewing and sharing/tweeting history to estimate the HIP parameters.

## 3.2 The popularity scale

It is well-known that network measurements such as the number of views and shares follow a long-tailed distribution. We quantify video popularity on an explicit popularity percentile scale, with 0.0% being the least popular, and 100% being the most popular. Fig. 3(a) and (b) show the popular-ity scale as boxplots (in log-scale) over the Active dataset, after 60 days of video life for shares and views, respectively. The shape of the scale is similar for both shares and views, and it reflects their long tail distribution. The only notable difference is the scale of the y-axis, as videos tend to accumulate less shares than views. The popularity scale for tweets is very similar to the one for shares, and shown in the appendix [1]. Based on the shares and views popularity scales, we define two mapping functions $S_t(x), P_t(x) : \mathbb{R}_+ \longrightarrow [0,1]$. Each function takes an argument – the number of shares for $S_t(x)$ or the number of views for $P_t(x)$ – and outputs the percentile value on the corresponding popularity scale constructed at time $t$.

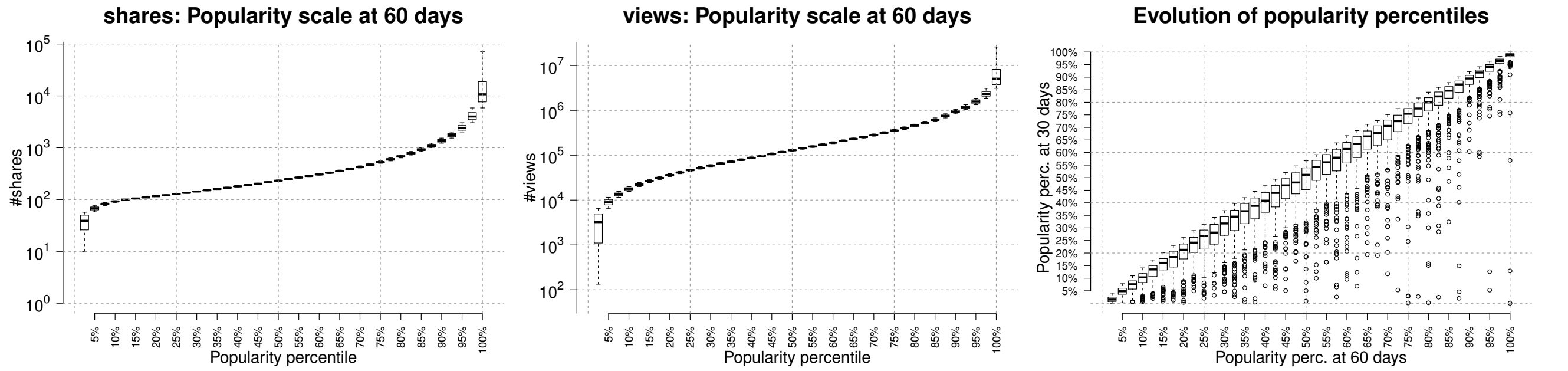

(a) (b) (c)

Figure 3: The popularity scale of YouTube videos, computed on the Active dataset. The total numbers of `shares` (a) and `views` (b) obtained by each video in the first 60 days after upload are divided into 40 equally spaced bins (i.e. each with 2.5% of the videos). Boxplots of shares/views in each bin are shown. The 2.5% most popular videos span more than one order of magnitude for both views and shares. Note that outliers in this bin are not represented, as the most popular videos in the collection have $\sim 10^8$ views and $\sim 10^6$ shares. (c) Evolution of the views popularity between 30 (y-axis) and 60 (x-axis) days. Boxplots show where each 2.5% of videos at 60 days came from (in terms of percentile position at 30 days). The outliers are videos that have improved their popularity significantly.

In Fig. 3(c) we explore the change of views popularity of each video from 30 days (y-axis) to 60 days (x-axis). Formally, we plot the relation between $P_{30}\left(\sum_1^{30} \bar{\xi}[t]\right)$ and $P_{60}\left(\sum_1^{60} \bar{\xi}[t]\right)$, where $\bar{\xi}[t]$ is the number of views at time $t$ (here the $t$-th day). Note that most videos retain a similar rank (in the boxes along the 45 degree diagonal line), or have a slight rank decrease as they are overtaken by other videos (slightly above the diagonal in the plot). No outliers exist in the upper-left part of the graph, since a video cannot lose viewcount that it already gained. Most notably, we can see that videos from any bin can *jump* to the top popularity bins between 30 and 60 days of age, such as the outliers for the few boxes on the far right. This phenomenon elicits two important questions: how did these videos go viral, and is this phenomenon related to external promotions?

# 4. THE ENDO-EXO MAP

Using two quantities defined in Sec 2.5, we construct a 2-dimensional map with *endogenous response* $A_{\hat{\xi}}$ as the x-axis and *exogenous sensitivity* $\mu$ as the y-axis. We call this plot the *endo-exo map*. This section presents example uses of this map for explaining video popularity, and identifying videos that are not promotable.

**Explaining popularity.** Intuitively, a video with a large endogenous response $A_{\hat{\xi}}$ and a high exogenous sensitivity $\mu$ has high potential to become viral. Specifically, each unit of exogenous excitation will generate $\mu A_{\hat{\xi}}$ events through the Hawkes intensity process. On the endo-exo map, videos in close proximity have similar potentials to become popular and the differences in their popularity would be due solely to the difference in exogenous attention. Fig 4(a) illustrates this phenomena using four videos. Videos $\text{v}_1$ and $\text{v}_2$ are very similar in both $A_{\hat{\xi}}$ and $\mu$; the fact that $\text{v}_1$ has 4.61x more views is explained by it receiving 3.22x more exogenous promotions. On the same map, $\text{v}_4$ received a similar amount of promotion as $\text{v}_1$ and their differences in popularity are explained by $\text{v}_4$ being less endogenously responsive (smaller $A_{\hat{\xi}}$) than $\text{v}_1$. Moreover, $\text{v}_3$ has a similar endogenous response and sees similar amounts of promotion as $\text{v}_1$; the difference between their popularities is explained by $\text{v}_3$ being less exogenously sensitive, with a lower $\mu$. The endo-exo map provides two distinct aspects from which a video's popularity can be analyzed, which are detailed next.

**What describes the most popular videos?** One may wonder whether higher popularity can be attributed to higher exogenous sensitivity, higher endogenous response or a combination of both. We examine a collection containing diverse video categories and find that the explanation varies. We draw on the endo-exo map all the videos that belong to the same category in the Active dataset and we visualize them as two-dimensional density plots. Fig. 4 (c) and (d) compares the plots for the videos in `Gaming` and `Film & Animation`, to that of the top 5% most popular videos in these two categories, respectively. Visibly, while most popular videos in `Film & Animation` are described by higher exogenous sensitivity (shifting upwards), the most popular `Gaming` videos have higher endogenous response – their density mass is shifted to the right of the endo-exo map. Other categories such as `Comedy` or `News & Politics` (shown in the appendix [1]) present two dense regions, one for higher $A_{\hat{\xi}}$ and one for higher $\mu$. These observations show that the most popular videos in different categories differ in terms of the two main factors that drive popularity.

**Identifying unpromotable videos.** The endo-exo map can be used to readily identify an interesting class of videos: the ones which are very difficult to promote. Given that the quantity $\mu A_{\hat{\xi}}$ describes the number of views that one unit of external promotion (via sharing or tweeting) will generate under the joint influence of endo- and exo- factors – a very small $\mu A_{\hat{\xi}}$ (e.g., $\mu A_{\hat{\xi}} < 1e-3$) is a hallmark of a video being *unpromotable*. Fig. 4(b) contains a zoomed-out view of the endo-exo map associated with the category `People & Blogs`. We found 63 videos ($\sim$ 3.2%) in this category to be unpromotable. Overall, 549 ($\sim$ 3.9%) videos in the Active set are deemed unpromotable. The thumbnail of one example video (a teenager video blog) is shown. It has $\mu = 2.88 \times 10^{-15}$ and $A_{\hat{\xi}} = 1$, hence each online promotion is expected to generate 0 views. In contrast, for video $v_1$ in Fig. 4(a), each promotion is expected to generate 598 views.

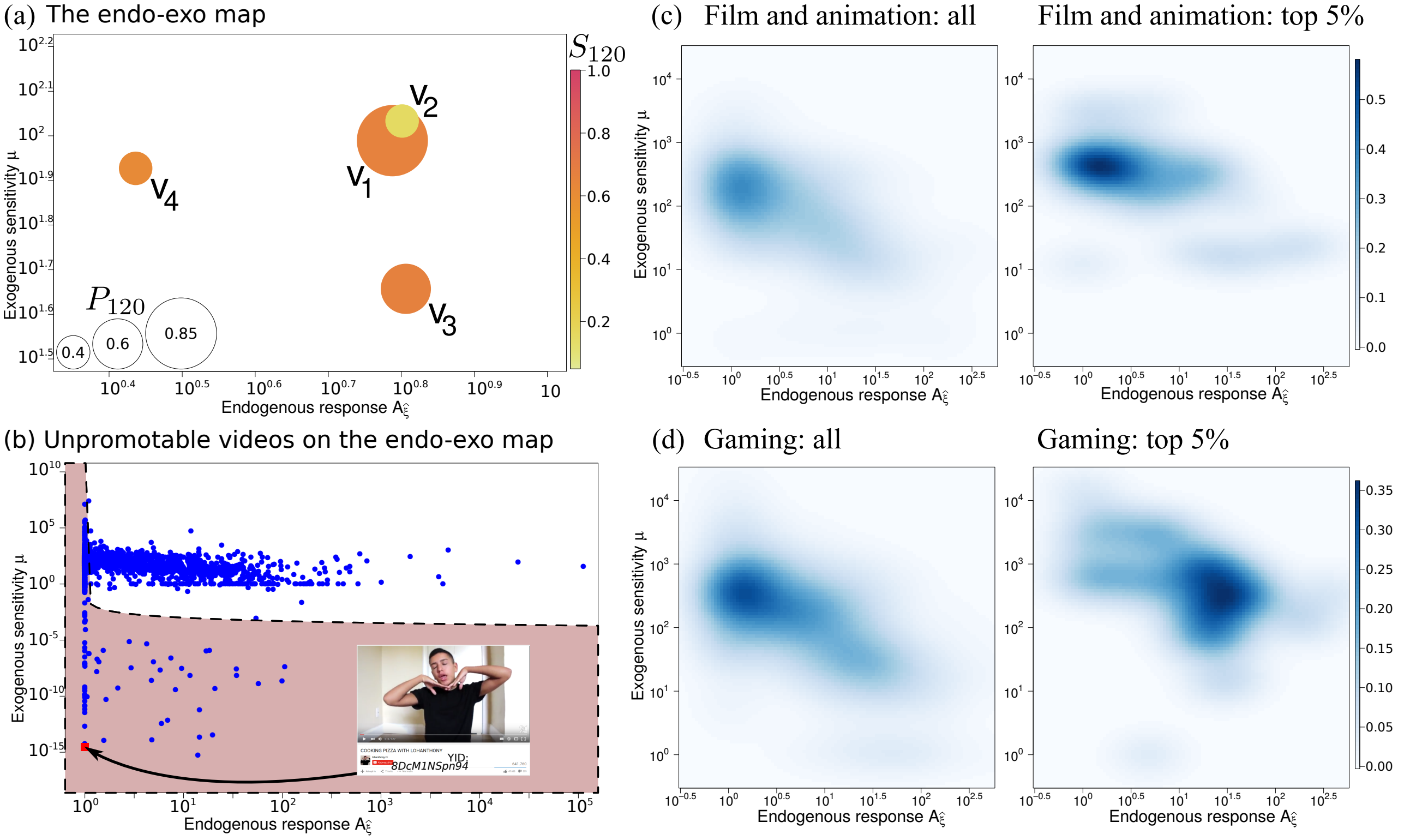

Figure 4: Visualizing video virality and video popularity using the *endo-exo map*. **(a)** Four example videos on the endo-exo map. X-axis $A_{\hat{\xi}}$: the magnitude of endogenous reaction; Y-axis $\mu$: sensitivity to exogenous stimuli. The radius of each circle is proportional to the *popularity percentile* $P_t(\cdot)$ of each video after $t = 120$ days, with values between 0.0 (least popular) and 1.0 (most popular). The color represents the amount (percentile) of total shares received, denoted as $S_t(\cdot)$, with values between 0.0 (no promotion) and 1.0 (receiving the most promotions). $v_1$ and $v_2$ present similar endogenous reaction and exogenous sensitivity, being at the same position on the endo-exo map. The difference in their popularity (size) is explained by the fact that $v_1$ received 3.22 times more promotions than $v_2$. Both $v_3$ and $v_4$ receive similar amounts of promotion (color) as $v_1$, but they achieve lower popularity (smaller size) due to their less privileged position on the endo-exo map: $v_3$ is less sensitive to external stimuli than $v_1$ and $v_2$, while $v_4$ has a smaller endogenous reaction than $v_1$ and $v_2$. Information about the four example videos are as follows, with their popularity percentile $P_{120}$ and shares percentile $S_{120}$: $v_1$ is a short `Gaming` video, YoutubeID *0lTTWeavl1c*, $P_{120}(634{,}370 \text{ views}) = 85\%$, $S_{120}(351 \text{ shares} = 65\%$; $v_2$ is a collection of "ALS ice bucket challenge" videos, YoutubeID *3hSIh-tbiKE*, $P_{120}(137{,}481) = 40\%$, $S_{120}(109) = 10\%$; $v_3$ is a funny science video, explaining types of infinity in math, YoutubeID *23I5GS4JiDg*, $P_{120}(193{,}052) = 60\%$, $S_{120}(356) = 65\%$; $v_4$ is from a Portuguese youtuber, YoutubeID *0ndmJzEIcgU*, $P_{120}(93{,}959) = 40\%$, $S_{120}(311) = 60\%$. **(b)** A zoomed-out scatter plot of the endo-exo map of the videos in the `People & Blogs` category. The shaded portion of this map consists of videos with low values of total response $\mu A_{\hat{\xi}} < 10^{-3}$ and hence dubbed *unpromotable* videos. Thumbnail of an example video *8DcM1NSpn94* is included, with $\mu = 2.88 \times 10^{-15}$ and $A_{\hat{\xi}} = 1$. **(c)** Density plot for all (left) vs the most popular 5% (right) Film & Animation videos. **(d)** Density plot for all (left) vs the most popular 5% (right) Gaming videos. Popular `Film and Animation` videos tend to have a higher exogenous sensitivity, while those for `Gaming` have mainly a higher endogenous response.

## 5. FORECASTING POPULARITY GROWTH

Via the endo-exo map, the Hawkes intensity process prescribes a video's expected popularity dynamics under external promotions. This section explores the predictive power of such a model. We first illustrate the setting for popularity forecasts using video examples, and then present a quantitative evaluation.

### 5.1 A video that will go viral

We use HIP to identify videos that are not already popular but have a high potential to become so. This is similar to the phenomenon of delayed recognition in science [21]. Note that this approach is predictive in that we aim to find such potentially viral items before they become popular, rather than a measurement-driven approach that analyzes viral items in past history. Video *1PuvXpv0yDM* in Fig. 5(a) is such an example, it received 15,687 views after being online for 90 days. The HIP model deems it to have a high endogenous response ($A_{\hat{\xi}} = 6.94 \times 10^{72}$) and a high exogenous sensitivity ($\mu = 119.02$). Between days 91 and 120, the video received an additional 229 shares, more than 6 times the number of shares during its first 90 days. Consequently, the video gained 2.42 million views, drastically improving its ranking on the popularity percentile scale from 5.85% to 94.9%.

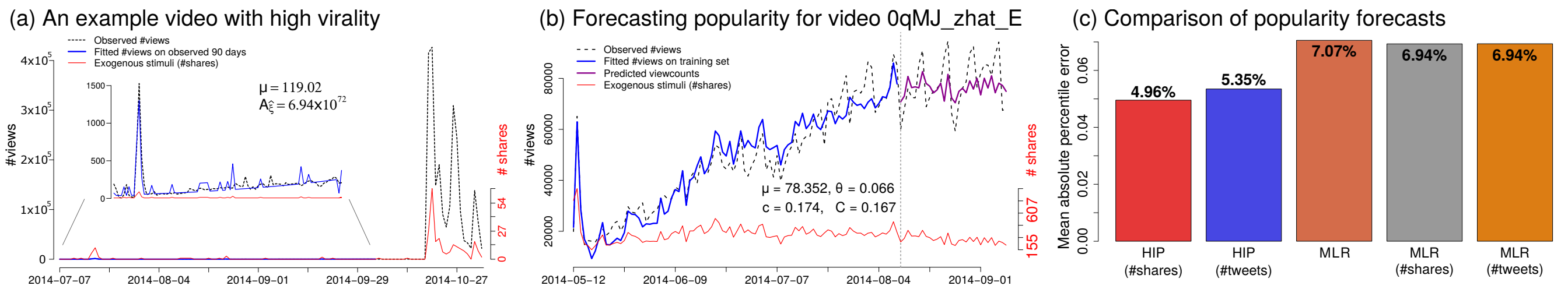

Figure 5: Popularity forecasting using the Hawkes intensity process. (a) Popularity series for video *1PuvXpv0yDM*, explaining a brain disorder. The video receives a total of 36 shares and 15,687 views in the first 90 days (see inset), it is estimated to have a high exogenous sensitivity and a high endogenous response ($\mu = 119.02$, $A_{\hat{\xi}} = 6.94 \times 10^{72}$). Between day 91 and day 120, this video *jumped* from a popularity percentile of 5.85% to 94.9%, receiving 229 shares and gaining 2.42 million views. (b) Forecasting popularity for video *0qMJ_zhat_E*. Black dotted line: viewcounts series from day 1 to 120 after video upload. Red line: exogenous simuli $s(t)$, also used in parameters estimation. Left of the the gray dashed vertical line at $T \leq 90$ days: time period used for parameters estimation. Blue line: fitted viewcounts for $T \leq 90$ days, generated using Eq 3. Magenta line: viewcount forecast for day 91 to 120. (c) Comparison of average forecasting errors on the ACTIVE set. y-axis: Forecasting errors, calculated as the absolute difference between the popularity percentile at day 120 and that forecasted by each approach. x-axis, left to right: Hawkes intensity model, using either `#shares` or `#tweets` as $s(t)$; multivariate linear regression (MLR), using only popularity history, or `#shares` and `#tweets`, respectively.

## 5.2 Evaluation of forecast

The HIP model takes as input the exogenous promotion $s[t]$ to produce estimates of the viewcount $\xi[t]$. To construct $\xi[t]$ in the future, $s[t]$ needs to be either estimable or known. We call this *forecasting* popularity, as opposed to predicting popularity where no information about future exogenous stimuli is assumed. Forecasting popularity has broad applications, such as estimating the effect of intended (promotional) interventions, and making decisions about when to promote.

**Evaluating popularity forecast on temporal hold-out data.** We design a protocol to quantitatively evaluate the predictive power of HIP. We use historical data held-out over time, thus avoiding the practical difficulty of generating realistic promotions and responses in a large-scale social network. Using (known) exogenous promotion $s[t]$, we forecast the popularity $\sum_{91}^{120} \xi[t]$ during the evaluation period (in purple) using Eq (4). Fig. 5(b) illustrates this setting with an example music video. A vertical line divides the observation period, day 1 to 90, and the evaluation period, day 91 to 120. The viewcount and the sharing history in the observation period is used to fit model parameters and explain observed popularity (in blue). For this example, the forecast and the actual views are fairly similar.

**Percentile-error metric.** We obtain a predicted total viewcount over the evaluation period, i.e, $\sum_{91}^{120} \xi[t]$, and we evaluate the performances by comparing it to the actual total viewcount $\sum_{91}^{120} \bar{\xi}[t]$. Commonly-used performance error metrics, such as root-mean-square-error (RMSE) or the normalized RMSE, are skewed by the large number of outliers in a long-tailed viewcount distribution and we chose not to use them. Instead, we map the forecasted number views to a popularity scale constructed as shown in Sec. 3.2, on the period 91-120 days of video life. We normalize the number of views into a metric between 0 and 1 and we compute the absolute error of the predicted percentile. When compared to the error metrics based on the difference in views (like RMSE), this metric focuses on ranking videos correctly with respect to a large collection and is as useful as the broad class of learning to rank applications.

**Baseline algorithms.** The state-of-the-art approach for popularity prediction uses multivariate linear regression (MLR), based on the observation that historic viewcounts are predictive for future viewcounts [30, 34]. We train linear regressors to predict daily viewcounts for each day between 91 and 120, using a 90-dimensional feature corresponding to the number of views in days 1 to 90. To give the MLR forecast the same amount of information as the HIP model, we build two enhanced baselines, denoted by MLR (#shares) and MLR (#tweets), by introducing the exogenous influence as additional variables, both in the training and in the prediction. Note that the HIP models for each video are learned and evaluated independently, all baselines are trained on ACTIVE and we obtain predictions for each video using cross-validation.

## 5.3 Forecasting results

Fig. 5(c) summarizes forecasting performance for HIP and the MLR baselines. The forecasts made using HIP have lower average error compared to the linear regression with or without exogenous stimuli (#shares, #tweets). The best forecast obtained an average percentile error of 4.96% (median 3%) for HIP (#shares) and 6.94% (median 3.75%) for the MLR (#shares), corresponding to a 28.6% relative reduction of error. These differences are statistically significant with paired t-test $p < 0.001$, and with a medium effect size according to Cohen's $d$ [12]. Within the HIP variants, we found that using the number of shares generates slightly better forecast than the number of tweets, but the differences are not statistically significant at $p = 0.001$ (more details about effect sizes and statistical tests can be found in the online appendix [1]). We speculate that the difference in forecasting performance is due to the nature of the sources of exogenous excitation: shares capture the promotion behavior via a multitude of environments, whereas tweets count the volume of promotion in Twitter only.

We also observe that the performance gap doubles when forecasting popularity on more difficult videos – videos with a large exogenous shock in the forecasting period, defined as the mean number plus 100 times the standard deviation of the number of shares during the observed period. Fig. 5(a) shows an example of such a video. There are 4006 such videos in the Active dataset, for which HIP (#shares) achieves a mean percentile error of 5.11% (median 3.25%), whereas MLR (#shares) achieves a mean error of 9.24% (median 6.5%). A typical situation when HIP misses the forecast is when none or very little external influence is recorded during the observed period and during which the popularity is likely to have been driven by unseen exogenous sources.

Lastly, a note about causality: HIP is linear control system with feedback loop, it is *causal* in a linear system sense [29] in that future tweets cannot change past views, but does not directly correspond to the causal inference paradigm about whether a control variable will change a response variable in the presence of other confounding factors. Nonetheless, we conducted statistical tests using the well-known Granger Causality [18] on the shares and view series (details in the appendix [1]); they do not show consistent results for either shares influencing views or vice versa.

# 6. RELATED WORK

**Popularity modeling and prediction.** Early measurement studies linked popularity with user influence in Twitter [7, 36] and with the speed and spread of information in social networks [8]. More recently, generative methods, usually based on point-processes, were introduced for popularity modeling [13, 15, 38] and prediction [4, 28]. In their seminal work, Crane and Sornette [13] showed how a Hawkes point-process can account for popularity bursts and decays. Subsequently, more sophisticated models have been proposed to model and simulate popularity in microblogs [38] and videos [15], by accounting for phenomena such as the "rich-get-richer" phenomenon and social contagion. Shen *et al.* [33] employ reinforced Poisson processes, modeling three phenomena: fitness of an item, a temporal relaxation function and a reinforcement mechanism. Zhao *et al.* [39] propose SEISMIC, which employs a double stochastic process, one accounting for infectiousness and the other one for the arrival time of events. TiDeH [23] is an extension of SEISMIC, which aims at estimating future number of views as a function of time, instead of just the final total cascade size. HIP differs from the above applications in two fundamental ways. First, most of the models [4, 23, 28, 31, 39] deal with single diffusion cascades, that is the reaction to single shocks. HIP models popularity as a continuous endogenous-exogenous intertwining, allowing it to closely fit complex evolutions. Second, typical point-process based methods require to observe each individual event during the training period, whereas HIP models volumes of attention directly.

**Modeling volumes of popularity.** A number of models have been proposed to describe the shape and evolution of the volume of social media activity over time. The seminal meme-tracker [24] system uses a curve with polynomial increase followed by exponential decay to describe sawtooth-shaped volume of news mentions. The SpikeM [27] system uses a fixed memory component, modulated by a periodic component, however it does not explicitly account for external influence. Most recently, Tsytsarau *et al.* [35] model the popularity volume as the convolutions two sequences, news event importance and media response, which are assumed to have predefined shapes. Yang *et al.* [37] propose a generative model to describe sequences that have multiple progression stages along with algorithms to estimate model parameters and to segment existing sequences. Being based a self-excited Hawkes process, HIP simultaneously addresses a series of shortcomings of the above approaches: it is adapted to forecast total popularity, it can recover all parameters from data, and it explains additional, non-stationary variations from linked data sources of external activities.

**Influence estimation and maximization** are somewhat related research problems, but distinct from the one approached in this paper. Influence estimation [17] aims to learn probabilities of influence between pairs of users, starting from a social graph and a log of actions of its users. Influence maximization [16, 22, 32] finds the subset of users who, if convinced to promote a piece of content, would maximize its diffusion. The main difference between this line of work and HIP is that we measure the volume of promotion and use it to forecast popularity, rather than taking a graph-centric view based on network structure and user interactions.

# 7. SUMMARY AND DISCUSSION

This research establishes a novel mathematical model to systematically link the endogenous response to the exogenous stimuli of a social system. The model developed here provides a nuanced view of the continued interactions of endogenous and exogenous effects that generate complex and multi-phased popularity dynamics over time. We validate the model on the popularity and promotion history of a large set of YouTube videos. We quantify the endogenous virality and exogenous sensitivity for each video, and we them to explain the properties of the most popular videos, as well as to identify videos that will respond well to promotions and those that will not. Such detailed analysis is possible because the aggregated attention and promotion data are available from YouTube or inferred from public sources such as Twitter. Note however that HIP does not make any platform-dependent assumption and that it can function with any popularity and promotion series generated by aggregated human behavior. We envision that the same kind of attention dynamics would hold for other content types, such as webpage views, podcasts, or blogs.

There are a number of simplifying assumptions and limitations of the proposed model, which can become fruitful directions of further investigation. The Hawkes intensity process captures popularity dynamics that are reflected only in the observed external promotion series, and does not capture other factors such as (daily or weekly) seasonality. What this model also focuses on is the *expected* influence over all users rather than individual influence. Both of these observations suggest extensions that could incorporate seasonality components as well as taking into account individual influences. Lastly, media items are influenced by a variety of sources in the open world and there are many sources of online or offline promotion that are unobserved or difficult to obtain data from. A well-known example is that gaming videos are known to be discussed intensively in topic-specific forums. Tracking and estimating diverse or even unknown sources of exogenous influence is another open research question.

**Acknowledgments.** This material is based on research sponsored by the Air Force Research Laboratory, under agreement number FA2386-15-1-4018. We thank the National Computational Infrastructure (NCI) for providing computational resources, supported by the Australian Government. We thank Alban Grastien, Richard Nock and Christian Walder for insightful discussions.

# **Appendix:** Expecting to be HIP: Hawkes Intensity Processes for Social Media Popularity

Marian-Andrei Rizoiu, Lexing Xie, Scott Sanner,
Manuel Cebrian, Honglin Yu, Pascal Van Hentenryck

## Contents

# 1 Details of HIP

Given time $t \in [0, \infty)$, we denote by $\lambda(t)$ the *event rate* of an online resource at time $t$. The goal of this section is to derive the *expected event rate*, denoted as $\xi(t)$, as the average response rate from a large network.

There are two sources of events in the social system – *exogenous* events originating outside the system and *endogenous* events spawned from within the system as the response to previous events (that are either exogenous or endogenous). For example, a public speech held by a famous politician can be an exogenous source for the number of views of relevant Youtube videos on politics; on the other hand, the views on trailers prior to the release of new movies exhibits a rich-get-richer effect for attention distribution that are characteristic of endogenous word-of-mouth diffusion.

## 1.1 Event rate in marked Hawkes processes

The Hawkes process [5], as defined in main text Eq. (1), is a non-homogeneous Poisson process with self-excitation, its event rate $\lambda(t)$, or instantaneous conditional intensity $r(t|\mathcal{H}_t)$ is:

$$\lambda(t) := r(t|\mathcal{H}_t) = \mu s(t) + \sum_{t_i < t} \phi_{m_i}(t - t_i) \tag{11}$$

Here $t > 0$ denotes time; $\mathcal{H}_t = \{(m_i, t_i); t_i < t\}$ is the event history before time $t$; $s(t)$ is the rate of exogenous events. $\mu$ is a scaling factor for exogenous stimulus, and $\phi_m(\tau)$ is a time-decaying trigger kernel that is determined by the magnitude of each event $m_i$.

Our Hawkes intensity model is a type of *marked Hawkes process*, as also used in [6, 12]. In marked Hawkes processes, each event $i$ has an occurrence time $t_i$ and a magnitude $m_i$ (or mark), which captures the relative amplification of the likelihood of an event to spawn future events. Intuitively, the mark can represent the magnitude of an earthquake when modeling the occurrence of aftershocks of an earthquake, or the number of people an event can subsequently influence when modeling social networks. This can be approximated, say, with the number of followers (on Twitter) or friends (on Facebook). The triggering kernel $\phi_m(\tau)$, as defined in main text Eq. (2), can be written as the product of two separable terms: $b(m)$ modeling the influence of event marks and $\phi(\tau)$ modeling the temporal decay:

$$\phi_m(\tau) = \kappa m^\beta \hat{\tau}^{-(1+\theta)} := b(m)\phi(\tau) \ ,$$
$$\text{with } b(m) = \kappa m^\beta, \phi(\tau) = \hat{\tau}^{-(1+\theta)}, \hat{\tau} = \tau + c. \tag{12}$$

Note $\hat{\tau} = \tau + c$ introduced for the sake of brevity in the calculations in the following sections.

In this work, we focus on modeling the total attention – commonly known as popularity, that closely connects to the average event rate across the whole network. We extract the number of followers $m$ for a large sample of users from our dataset described in Sec. 3 and fit a power-law distribution $p(m) = (\alpha-1)m^{-\alpha}$ following the method in [1]. We obtain $\alpha = 2.016$ and we use it throughout the experiments. As noted in the main text, the two power law exponents are distinct in meaning and function, $\theta$ defines memory decay over time, while $\alpha$ is determined by the user distribution at large. $\alpha$ is estimated from a large Twitter sample. $\theta$ and other video-dependent parameters are estimated from popularity history as detailed in Sec 2 below.

## 1.2 Proof of Theorem 2.1: $\xi(t)$ for unobserved point processes

In this section we give the proof of the main text Theorem 2.1. More precisely, we derive the *expected event rate* $\xi(t)$ over time, specified in the main text Eq. (3). This is done in three steps: we first include a preliminary description of the event rate $\lambda(t)$ in terms of the underlying counting process over infinitesimal intervals, we then derive the expected event rate for *unmarked* Hawkes processes, and finally we build upon these to derive the expected event rate for *marked* Hawkes processes.

### 1.2.1 Preliminaries: event rate, counting process

It is well known in stochastic process literature [8] that the event rate $\lambda(t)$, or the *conditional intensity specification* $r(t|\mathcal{H}_t)$ of a point process is completely characterized by the corresponding counting process $N(t)$. Here $N(t)$ is the total number of events observed between time 0 and $t$.

Given an infinitesimal interval $\delta$ at time $t$, the relationship between $N(t)$ and $r(t|\mathcal{H}_t)$ is described as:

$$\begin{aligned} &\mathbb{P}\left(N(t+\delta)-N(t)=1|\mathcal{H}_t\right) = r(t|\mathcal{H}_t)\delta + o(\delta) \ , \\ &\mathbb{P}\left(N(t+\delta)-N(t)>1|\mathcal{H}_t\right) = o(\delta) \ , \\ &\qquad \text{with } \lim_{\delta\downarrow 0}\frac{o(\delta)}{\delta} = 0 \ . \end{aligned} \quad (13)$$

Here $\mathbb{P}$ denotes the probability of a discrete random variable. The intuition of the expression above is that $r(t|\mathcal{H}_t)$ is proportional to the probability that $N(t)$ increments by 1, and that it is "very unlikely" for $N(t)$ to increment by more than one.

Let $dN_t$ be the counting increment $N(t+\delta) - N(t)$ as $\delta \downarrow 0$. From Eq. (13), we can describe $dN_t$ as a Bernoulli random variable, with:

$$\begin{aligned} \mathbb{P}(dN_t = 1|\mathcal{H}_t) &= r(t|\mathcal{H}_t)\delta \ , \\ \mathbb{P}(dN_t = 0|\mathcal{H}_t) &= 1 - r(t|\mathcal{H}_t)\delta \ , \\ &\text{for } \delta \downarrow 0 \ . \end{aligned}$$

It follows from the above that

$$E_{dN_t|\mathcal{H}_t}[dN_t] = r(t|\mathcal{H}_t)\delta, \text{ for } \delta \downarrow 0.$$

Using the shorthand $\lambda(t)$ for event rate and putting the above together, we can see that Hawkes processes can be specified as:

$$\begin{aligned} \lambda(t) := r(t|\mathcal{H}_t) &= \lim_{\delta\downarrow 0}\frac{\mathbb{P}\left(N(t+\delta)-N(t)=1|\mathcal{H}_t\right)}{\delta} \\ &= \lim_{\delta\downarrow 0}\frac{\mathbb{P}\left(dN_t=1|\mathcal{H}_t\right)}{\delta} \\ &= \lim_{\delta\downarrow 0}\frac{\mathbb{E}_{dN_t|\mathcal{H}_t}\left[dN_t\right]}{\delta} \ , \end{aligned} \quad (14)$$

Note that Eq. (14) is an alternate formulation of Eq. (11) through the counting process $N(t)$. Eq. (14) holds for all non-homogeneous Poisson processes. Hawkes processes (marked and unmarked) are special cases of non-homogeneous Poisson processes.

### 1.2.2 Expected event rate for *unmarked* Hawkes

We first study the simpler case of an *unmarked* Hawkes processes $\lambda^u(t)$, and derive its expected event rate $\xi^u(t)$ over possible event histories. While it is not strictly necessary to breakdown the derivation into two parts, this helps illustrate the main ideas underlying the derivation for *marked* processes in the next subsection. The key idea in this subsection is converting the conditional expectation of event history into increments of the counting process, and using conditional expectations to link the expectations of counting increments to the expected rate $\xi^u(t)$ via $\lambda^u(t)$. The next subsection will use exactly the same treatment for the history of event times, and performs a similar treatment for a history of event magnitudes.

Let an unmarked Hawkes process be:

$$\lambda^u(t) := r(t|\mathcal{H}_t) = \mu s(t) + \sum_{t_i<t}\phi(t-t_i). \quad (15)$$

Here $\phi(\tau)$ is a memory kernel specified in Eq. (12), scaling constant $\kappa$ is omitted without loss of generality. Note Eq. (14) still holds. Here $i = 1, \dots, N(t)$ is the event index, and $N(t)$ is a random variable, representing the total number of events before time $t$, i.e. the counting process. It is worth noting that there are two equivalent expressions of the event history $\mathcal{H}_t$. The first one is $\mathcal{H}_t$ being a *random* set of time stamps of the events which took place between $[0, t)$, $\mathcal{H}_t = \{t_{1:N(t)}\}$. Note that the cardinality of $\mathcal{H}_t$ is $N(t)$, hence random. The second definition expresses $\mathcal{H}_t$ with the counting process $N(\tau)$ as a piece-wise constant function between $[0, t)$. Here each jump point in $N(\tau)$ correspond to an event time, and the number of jumps is random variable $N(t)$. It is easy to see that the two definitions are equivalent. For convenience, we write $\mathcal{H}_t = \{N(\tau), 0 < \tau < t\}$.

We define the expected event rate $\xi^u(t)$ as a function over time, obtained by taking expectations of $\lambda^u(t)$ over the event history, note that this is an *unmarked* process, all event magnitudes are the same (i.e. 1). Note that taking expectations over $\mathcal{H}_t$ can be thought of as either over a set of random variables $t_i$ with a random dimensionality $N(t)$, or as over random piece-wise constant functions over time, i.e. $N(\tau)$, for $0 < \tau < t$.

$$\begin{aligned}
\xi^u(t) :=& \mathbb{E}_{\mathcal{H}_t}\left[\lambda^u(t)\right] \\
=& \mathbb{E}_{\mathcal{H}_t}\left[\mu s(t) + \sum_{t_i<t}\phi(t-t_i)\right] \\
\overset{6(a)}{=}& \mu s(t) + \mathbb{E}_{t_{1:N(t)}}\left[\sum_{i=1}^{N(t)}\phi(t-t_i)\right] \\
\overset{6(b)}{=}& \mu s(t) + \\
&+ \mathbb{E}_{\{N(\tau),0<\tau<t\}}\left[\lim_{\delta\downarrow 0}\sum_{k=1}^{K}\mathbf{1}(dN_{k\delta}=1)\phi(t-k\delta)\right] \\
\overset{6(c)}{=}& \mu s(t) + \\
&+ \lim_{\delta\downarrow 0}\mathbb{E}_{dN_{k\delta},k=1:K}\left[\sum_{k=1}^{K}\mathbf{1}(dN_{k\delta}=1)\phi(t-k\delta)\right]
\end{aligned} \tag{16}$$

Here step 6(a) is due to $\mu s(t)$ being not random, and that expectations over all $t_{1:N(t)}$ are equivalent to taking expectations over event history $\mathcal{H}_t$.

In step 6(b), we divide time interval $(0, t)$ into $K$ equal-sized infinitesimal intervals of size *delta*, with $K\delta = t$. $\mathbf{1}(dN_{k\delta} = 1)$ is the indicator function that takes value 1 when there is an event in the interval $[(k-1)\delta, k\delta)$, or $dN_{k\delta} = 1$, and 0 otherwise. Note that we replaced each arrival time $t_i$ from line 6(a) with $k\delta$, since event $i$ occurred in the time interval $[(k-1)\delta, k\delta)$. Consequently, the term $\phi(t-t_i)$ became $\phi(t-k\delta)$.

In step 6(c), we first exchange the order of the limit $\lim_{\delta\downarrow 0}$ and expectation. We note that taking expectation over the counting process $\{N(\tau), 0 < \tau < t\}$ is equivalent to taking expectation over its Bernoulli increments $\{dN\tau, 0 < \tau < t\}$, or it's discretized version over infinitesimal intervals $dN_{k\delta},_{k=1:K}$.

We then unroll the sum over all intervals. For an interval $k'\delta$, or $[(k'-1)\delta, k'\delta)$, the indicator function becomes $\mathbf{1}(dN_{k'\delta} = 1)$, the kernel $\phi(t-k'\delta)$, and the expectation is taken over $dN_{k\delta},_{k=1:k'}$ since the process is causal – i.e. current events are only influenced by the past and not the future.

$$\begin{aligned}
&\mathbb{E}_{dN_{k\delta},k=1:K}\left[\sum_{k=1}^{K}\mathbf{1}(dN_{k\delta}=1)\phi(t-k\delta)\right] \\
&\quad= \mathbb{E}_{dN_{k\delta},k=1}\left[\mathbf{1}(dN_{1\delta}=1)\phi(t-1\delta)\right] \\
&\qquad+ \mathbb{E}_{dN_{k\delta},k=1,2}\left[\mathbf{1}(dN_{2\delta}=1)\phi(t-2\delta)\right] \\
&\qquad+ \ldots \\
&\qquad+ \mathbb{E}_{dN_{k\delta},k=1:k'}\left[\mathbf{1}(dN_{k'\delta}=1)\phi(t-k'\delta)\right] \\
&\qquad+ \ldots \\
&\qquad+ \mathbb{E}_{dN_{k\delta},k=1:K}\left[\mathbf{1}(dN_{K\delta}=1)\phi(t-K\delta)\right]
\end{aligned} \tag{17}$$

Each expectation term in Eq. (17) can be computed as follows.

$$\begin{aligned}
&\mathbb{E}_{dN_{k\delta},k=1:k'}\left[\mathbf{1}(dN_{k'\delta}=1)\phi(t-k'\delta)\right] \\
\overset{(8a)}{=}& \mathbb{E}_{\mathcal{H}_{(k'-1)\delta}}\mathbb{E}_{dN_{k'\delta}|\mathcal{H}_{(k'-1)\delta}}\left[\mathbf{1}(dN_{k'\delta}=1)\right]\phi(t-k'\delta) \\
\overset{(8b)}{=}& \delta\mathbb{E}_{\mathcal{H}_{(k'-1)\delta}}\left[\lambda^u(k'\delta)\right]\phi(t-k'\delta) \\
\overset{(8c)}{=}& \delta\xi^u(k'\delta)\phi(t-k'\delta)\ ;
\end{aligned} \tag{18}$$

Step (18a) is due to the kernels term $\phi(t-k'\delta)$ being non-random and thus becoming constants. Also note that each expectation $\mathbb{E}_{dN_{k\delta},k=1:k'}\left[\mathbf{1}(dN_{k'\delta}=1)\right]$ can be computed by breaking down the joint distribution $dN_{k\delta},_{k=1:k'}$ into the conditional distribution $dN_{k'\delta}|\mathcal{H}_{(k'-1)\delta}$ and the prior distribution over $\mathcal{H}_{(k'-1)\delta}$. We write the prior in terms of event history for notational convenience. Due to the two equivalent definitions of event history, taking the expectation over the history $E_{\mathcal{H}_{(k'-1)\delta}}$ is equivalent to taking the expectation over increments of the counting process $E_{dN_{k\delta},k=1:(k'-1)}$.

Step (18b) is due to Eq. (14), the inner expectation can be written as $\mathbb{E}_{dN_{k'\delta}|\mathcal{H}_{(k'-1)\delta}}\left[\mathbf{1}(dN_{k'\delta}=1)\right] = \delta\lambda^u(k'\delta)$ as $\delta\downarrow 0$.

Step (18c) is due to the definition of the expected event rate $\xi^u(t)$ in Eq. (16). The expectation of event rate $\lambda^u(k'\delta)$ over $\mathcal{H}_{(k'-1)\delta}$ becomes $\xi^u(k'\delta)$ as $\delta\downarrow 0$.

Applying the result of Eq. (18) back to Eq. (17), we get:

$$\begin{aligned}
&\mathbb{E}_{dN_{k\delta},k=1:K}\left[\sum_{k=1}^{K}\mathbf{1}(dN_{k\delta}=1)\phi(t-k\delta)\right] \\
&\qquad= \sum_{k=1}^{K}\delta\xi^u(k\delta)\phi(t-k\delta)
\end{aligned} \tag{19}$$

Applying Eq. (19) to the end of Eq. (16), and taking the limit $\delta\downarrow 0$, we have:

$$\begin{aligned}
\xi^u(t) &:= \mathbb{E}_{\mathcal{H}_t}\left[\lambda^u(t)\right] \\
&\overset{t=K\delta}{=} \mu s(t) + \lim_{\delta\downarrow 0}\sum_{k=1}^{K}\delta\xi^u(k\delta)\phi(t-k\delta) \\
&= \mu s(t) + \int_0^t \xi^u(\tau)\phi(t-\tau)d\tau
\end{aligned} \tag{20}$$

Performing a change of variable $\tau \leftarrow t-\tau$, we obtain the integral equation specifying the expected event rate for unmarked Hawkes process.

$$\xi^u(t) = \mu s(t) + \int_0^t \xi^u(t-\tau)\phi(\tau)d\tau \tag{21}$$

To the best of our knowledge, this definition of the intensity function, along with the derivation of its analytical form is new. The original paper by Hawkes [5] presents an integral equation of similar form, but it is for the covariance density and not the event intensity function.

### 1.2.3 Expected event rate for *marked* Hawkes

The *expected event rate function* $\xi(t)$ for a *marked* Hawkes process is defined as the expectation of the event rate function $\lambda(t)$ over the set of event times and magnitudes before time $t$. In this subsection we work with the event rate as specified in Eq. (11):

$$\lambda(t) := r(t|\mathcal{H}_t) = \mu s(t) + \sum_{t_i<t} \phi_{m_i}(t-t_i)$$

In this subsection, we augment the definition of the event history with the event magnitudes, i.e., $\mathcal{H}_t = \{(t_i, m_i)_{,i=1:N(t)}\}$, In other words, each event consists of a (random) jump time $t_i$ and a (random) event magnitude $m_i$, and there are $N(t)$ (another random quantity) such time-magnitude pairs before time $t$.

We assume that any event magnitude $m_i$ is drawn *i.i.d.* from the same power-law distribution $p(m) = (\alpha-1)m^{-\alpha}, m>0$, once the event time $t_i$ is determined. That is to say, for an event spawned through the endogenous process, the magnitude of the event is independent of the magnitude of its parent event.

We define the expected event rate $\xi(t)$ for the *marked* Hawkes processes $\lambda(t)$ as follows. Step (22a) below is due to $\mu s(t)$ being non-random.

$$\begin{aligned}
\xi(t) &:= \mathbb{E}_{\mathcal{H}_t}[\lambda(t)] \\
&= \mathbb{E}_{\mathcal{H}_t}\left[\mu s(t) + \sum_{t_i<t}\phi_{mi}(t-t_i)\right] \\
&\overset{Eq.12}{=} \mathbb{E}_{\mathcal{H}_t}\left[\mu s(t) + \sum_{t_i<t} b(m_i)\phi(t-t_i)\right] \\
&\overset{(12a)}{=} \mu s(t) + \mathbb{E}_{\mathcal{H}_t}\left[\sum_{i=1}^{N(t)} b(m_i)\phi(t-t_i)\right]
\end{aligned} \tag{22}$$

In order to unroll this expectation into $K$ small intervals of size $\delta$, we need a set of auxiliary variables, called $m_k, k=1:K$. For each interval $[(k-1)\delta, k\delta)$, we draw $m_k \sim p(m)$. If indicator function $\mathbf{1}(dN_{k\delta}=1)=1$, then $m_k$ is *kept*; otherwise when $\mathbf{1}(dN_{k\delta}=1)=0$, $m_k$ is *thrown away* as no event happened in this interval. One can easily verify that this process of generating *i.i.d.* draws of $m_k$ is equivalent to obtaining *i.i.d.* draws of the original $m_i$. We use this to re-write the expectation in Eq. (22), and exchange the order of the expectation and the limit.

$$\begin{aligned}
&\mathbb{E}_{\mathcal{H}_t}\left[\sum_{i=1}^{N(t)} b(m_i)\phi(t-t_i)\right] \\
=&\mathbb{E}_{\mathcal{H}_t}\left[\lim_{\delta\downarrow 0}\sum_{k=1}^{K}\mathbf{1}(dN_{k\delta}=1)b(m_k)\phi(t-k\delta)\right] \\
=&\lim_{\delta\downarrow 0}\mathbb{E}_{\mathcal{H}_t}\left[\sum_{k=1}^{K}\mathbf{1}(dN_{k\delta}=1)b(m_k)\phi(t-k\delta)\right]
\end{aligned} \tag{23}$$

We exchange the order of the expectation and the summation, and unroll the sum over all intervals. This is similar to Eq. (17). A marked Hawkes process respects event sequencing in time (also called *causal* in linear systems), this means that the expectation of each term $[\mathbf{1}(dN_{k\delta}=1)b(m_k)\phi(t-k\delta)]_{k=k'}$ only depends on event history $\mathcal{H}_{k'\delta}$, and not after.

$$\begin{aligned}
&\mathbb{E}_{\mathcal{H}_t}\left[\sum_{k=1}^{K}\mathbf{1}(dN_{k\delta}=1)b(m_k)\phi(t-k\delta)\right] \\
&=\mathbb{E}_{\mathcal{H}_{1\delta}}\left[\mathbf{1}(dN_{k\delta}=1)b(m_k)\phi(t-k\delta)\right]_{k=1} \\
&\quad+\mathbb{E}_{\mathcal{H}_{2\delta}}\left[\mathbf{1}(dN_{k\delta}=1)b(m_k)\phi(t-k\delta)\right]_{k=2} \\
&\quad+\dots \\
&\quad+\mathbb{E}_{\mathcal{H}_{k'\delta}}\left[\mathbf{1}(dN_{k\delta}=1)b(m_k)\phi(t-k\delta)\right]_{k=k'} \\
&\quad+\dots \\
&\quad+\mathbb{E}_{\mathcal{H}_{K\delta}}\left[\mathbf{1}(dN_{k\delta}=1)b(m_k)\phi(t-k\delta)\right]_{k=K}
\end{aligned} \tag{24}$$

We now compute the expectation term when $k=k'$. Step (15a) is due to $\phi(t-k'\delta)$ being non-random. Step (15b) breaks down the expectation over the entire history $\mathcal{H}_{k'\delta}$ into the part over the current event (and its magnitude if happens) $dN_{k'\delta}, m_{k'}$ conditioned on prior history $\mathcal{H}_{(k'-1)\delta}$ and over the prior history itself. This is similar to Eq. (18) for the unmarked process.

$$\begin{aligned}
&\mathbb{E}_{\mathcal{H}_{k'\delta}}\left[\mathbf{1}(dN_{k\delta}=1)b(m_k)\phi(t-k\delta)\right]_{k=k'} \\
\overset{(15a)}{=}&\mathbb{E}_{\mathcal{H}_{k'\delta}}\left[\mathbf{1}(dN_{k'\delta}=1)b(m_{k'})\right]\phi(t-k'\delta) \\
\overset{(15b)}{=}&\mathbb{E}_{\mathcal{H}_{(k'-1)\delta}}\mathbb{E}_{dN_{k'\delta},m_{k'}|\mathcal{H}_{(k'-1)\delta}} \\
&\left[\mathbf{1}(dN_{k'\delta}=1)b(m_{k'})\right]\phi(t-k'\delta)
\end{aligned} \tag{25}$$

Note that the middle conditional expectation term decomposes into two parts. Note that $\mathbf{1}(dN_{k'\delta}=1)$ is independent of $m_{k'}$, therefore

$$\mathbb{E}_{dN_{k'\delta}|\mathcal{H}_{(k'-1)\delta}}\mathbf{1}(dN_{k'\delta}=1) = \delta\lambda(k'\delta);\ \text{as } \delta\downarrow 0 \tag{26}$$

The expectation of function $b(m)$ is the same and only depends on $p(m)$ whenever $dN_{k'\delta}=1$. This is due to the generating assumption of event magnitudes at the beginning of this subsection.

$$\mathbb{E}_{dN_{k'\delta},m_{k'}|\mathcal{H}_{(k'-1)\delta}} b(m_{k'}) = \mathbb{E}_m[b(m)] \tag{27}$$

Furthermore, we see that $\mathbb{E}_m[b(m)]$ can be computed in closed form, we call this modeling constant $C$ (also defined in main text Theorem 2.1).

$$\mathbb{E}_m[b(m)] = \mathbb{E}_m[\kappa m^\beta] = \kappa \int_1^\infty p(m) m^\beta dm$$
$$= \kappa \int_1^\infty (\alpha - 1) m^{-\alpha} m^\beta dm = \frac{\kappa(\alpha-1)}{\alpha - \beta - 1} := C \tag{28}$$

Plugging in the result of Eq. (26)–28 back to Eq.25, we notice $\mathbb{E}_{\mathcal{H}_{(k'-1)\delta}}[\lambda(k'\delta)] = \xi(k'\delta)$ due to the definition of $\xi(t)$ in Eq. (22), which yields

$$\mathbb{E}_{\mathcal{H}_{k'\delta}} [\mathbf{1}(dN_{k\delta} = 1) b(m_k) \phi(t - k\delta)]_{k=k'} = C \cdot \delta\xi(k'\delta). \tag{29}$$

Applying this result to Eq. (24) and then to Eq.22, followed by taking the limit $\delta \downarrow 0$, we have:

$$\begin{aligned}
\xi(t) &:= \mathbb{E}_{\mathcal{H}_t}[\lambda(t)] \\
\overset{t=K\delta}{} &= \mu s(t) + \lim_{\delta\downarrow 0} \sum_{k=1}^{K} C \cdot \delta\xi(k\delta)\phi(t - k\delta) \\
&= \mu s(t) + C \int_0^t \xi(\tau)\phi(t-\tau) d\tau \\
\overset{\tau \leftarrow t-\tau}{} &= \mu s(t) + C \int_0^t \xi(t-\tau)\phi(\tau) d\tau \\
\overset{\phi(\tau)=\hat{\tau}^{-(1+\theta)}}{} &= \mu s(t) + C \int_0^t \xi(t-\tau)\hat{\tau}^{-(1+\theta)} d\tau
\end{aligned} \tag{30}$$

Eq. (30) is Eq. (3) in the main text Theorem 2.1.

### 1.3 Branching factor and endogenous response

We derive two quantities from the Hawkes Intensity Process in order to better visualize and explain the diverse behavior of video popularity.

**Branching factor $n$** The first key parameter is the branching factor $n$, defined as the mean number of daughter events generated by a mother event. For a marked Hawkes point process, the branching factor is computed by integrating the triggering kernel over time and taking the expectation over the magnitude $m$.

$$n = \int_{m_{min}}^\infty \int_0^\infty p(m)\phi_m(\tau) d\tau dm =$$
$$= \frac{C}{\theta c^\theta}, \text{ for } \beta < \alpha - 1 \text{ and } \theta > 0 \ .$$

$n < 1$ implies a *subcritical regime*, i.e., the instantaneous rate of events decreases over time and new events will eventually cease to occur (in probability); $n > 1$ implies a *supercritical regime*, i.e. each new event generates more than one direct descendant, which in turn generates more descendants, unless the network condition changes, the total number of events is expected to be infinity.

**Endogenous response $A_{\hat{\xi}}$** The second quantity $A_{\hat{\xi}}$, as defined in the main text Sec. 2.5, is the total number of (direct and indirect) descendants generated from one event. In the main text, $A_{\hat{\xi}}$ is defined in the discrete form, however it can also be defined as an integral over the continuous form of the impulse response as $A_{\hat{\xi}} = \int_0^\infty \hat{\xi}(t)dt$.

Although defined separately, we can see that $A_{\hat{\xi}}$ is closely related to branching factor $n$: the initial exogenous event will generate $n$ events as first-generation direct descendants. Each of these events will generate an expected $n$ events ($n^2$ events in the second generation), and each of these will in turn generate $n$ events ($n^3$ events in the third generation), .... Here $n^k$ is the average number of events in the $k^{th}$ generation, and so on. This leads to an equivalent definition of $A_{\hat{\xi}}$.

$$\begin{aligned}
A_{\hat{\xi}} &= 1 + n + n^2 + \ldots + n^k + \ldots = \\
&= \lim_{k\to\infty} \frac{1-n^k}{1-n} = \\
&= \begin{cases} \frac{1}{1-n} & , n < 1 \\ \infty & , n > 1 \end{cases}
\end{aligned} \tag{31}$$

While both capturing the endogenous property of the Hawkes Intensity model, $A_{\hat{\xi}}$ and $n$ emphasize different intuitions. We chose to visualize $A_{\hat{\xi}}$ in the endo-exo map, because it has a direct correspondence to the *sliced* LTI system view in main text Eq. (7) and Fig. 2(b), and that $A_{\hat{\xi}}$ has better numerical resolution for the more viral videos – i.e., when $n$ is close to 1. In the main text, we obtain estimates of $A_{\hat{\xi}}$ by numerically summing $\hat{\xi}[t]$ in the main text Eq. (7) over $10,000$ discrete time steps.

Crane and Sornette [3] showed that the Hawkes Intensity Process in a super-critical state could explain some rising patterns of popularity observed in social media. We note, however, that finite resources in the real world, such as collective human attention [10], are bound to be exhausted and online systems cannot stay indefinitely in a supercritical regime. We argue, most online media items are affected by a continued interaction of exogenous stimuli and endogenous reaction (that may be sub- or supercritical), leading to continued rise in popularity, or multiple phases of rising and falling patterns.

### 1.4 HIP as an LTI system

**Proof of main text Corollary 2.2** The Hawkes intensity process can be viewed as a system with one input – the exogenous stimuli rate $s(t)$, and one output – the event rate $\xi(t)$. The main text Corollary 2.2 states that the system $s(t) \to \xi(t)$ is an Linear Time Invariant (LTI) system. That is to say, the system has two properties:

*Linearity*, which states that the relation between the input and the output of the system is a linear map: if $s_1(t) \to \xi_1(t)$ and $s_2(t) \to \xi_2(t)$, then $as_1(t) + bs_2(t) \to a\xi_1(t) + b\xi_2(t), \forall a, b \in \mathbb{R}$. We can see that linear scaling is true $as_1(t) \rightarrow a\xi_1(t)$ by multiplying $a$ to both sides of Eq. (3) in the main text and re-grouping terms. Additivity $as_1(t) + bs_2(t) \rightarrow a\xi_1(t) + b\xi_2(t)$ can be shown similarly.

*Time invariance*, which states that the response to a delayed input is identical and similarly delayed: if $s(t) \rightarrow \xi(t)$ then $s(t - t_0) \rightarrow \xi(t - t_0)$.

We wish to show the following for Eq. (3) of the main text:

$$\xi(t - t_0) = \mu s(t - t_0) + C \int_0^t \hat{\tau}^{-(1+\theta)} \xi(t - t_0 \ - \tau) d\tau$$

After a change of variable $t' = t - t_0$, we can see that the LHS is $\xi(t')$. For the RHS, $\hat{\tau}$ remains unchanged, the rest is:

$$\mu s(t') + C \int_0^{t'+t_0} \hat{\tau}^{-(1+\theta)} \xi(t' - \tau) d\tau$$

We write the integral into two parts, i.e., $(0, t)$ and $(t', t' + t_0)$.

$$\mu s(t') + C \int_0^{t'} \hat{\tau}^{-(1+\theta)} \xi(t' - \tau) d\tau + C \int_{t'}^{t'+t_0} \hat{\tau}^{-(1+\theta)} \xi(t' - \tau) d\tau$$

We note that $\xi(t)$ is a causal function, i.e., $\xi(t) = 0$ for $t < 0$, or $\xi(t' - \tau) = 0$ for $\tau > t'$. The second term vanishes. RHS becomes

$$\mu s(t') + C \int_0^{t'} \hat{\tau}^{-(1+\theta)} \xi(t' - \tau) d\tau$$

Note LHS = RHS due to Eq. (30) and time invariance holds.

The main text Corollary 2.2 concerning the LTI property directly implies the following about Hawkes intensity processes, as illustrated in Fig. 2(b) of the main text.

- *Additive effects from multiple sources of external stimulation*: when applying two sources of excitation, the event rate of the resulting Hawkes intensity process is the sum of the rates generated by each source of excitation independently. This allows us to separately quantify the impact of each source.
- *Scaling the expected event rate*: if the exogenous stimuli scales up or down, the endogenous reaction will scale accordingly. In other words, if we can control the amount of exogenous promotions, we could boost or suppress the number of views for videos that respond to such promotions.
- *Shifting in time*: if the exogenous stimuli is shifted in time, so will the views responding to it. In other words, we could schedule promotions (and subsequent views) for videos that respond to such promotions.

**Proof of main text Lemma 2.3** The *sliced* fitting graph in Fig. 2(b) of the main text can be understood as an illustration of these three properties. In reality we observe the exogenous stimuli $s(t)$ as a discretized function (denote discrete time index as $[t]$) consisting a series of impulses located at $\tau = 1, 2, \ldots, T$, i.e.

$$\sum_{\tau=0}^{T} s[\tau]\delta[t - \tau] \tag{32}$$

Directly following from the three properties, we can see that $\xi[t]$ is a superposition of impulse response function $\hat{\xi}[t]$ scaled by $s[\tau]$ and shifted by the corresponding amount, i.e.

$$\xi[t] = \sum_{\tau=0}^{T} s[\tau]\hat{\xi}[t - \tau] \tag{33}$$

Visibly, Eq. (33) is Eq. 8 in the main text Lemma 2.3.

# 2 Details about fitting HIP

This section describes some of the implementation and computational details for estimating the model in Eq. (30) from observed popularity and promotion histories.

## 2.1 The loss function

In this section, we develop the calculation of the loss function defined in main text Eq. (6). For each video with observed $\{\bar{\xi}[t], \bar{s}[t], t = 1, \ldots, T\}$, we find an optimal set of models parameters $\{\mu, \theta, C, c\}$ and also estimate the unobserved external influence (parameters $\gamma$ and $\eta$). This is done by minimizing the square error between the series $\bar{\xi}[t]$ and the model $\xi[t]$, $\forall t \in 0, 1 \ldots T$. The corresponding optimization problem is as follows:

$$\begin{aligned} \min_{\mu,\theta,C,c,\gamma,\eta} J &= \frac{1}{2} \sum_{t=0}^{T} \left(\xi[t] - \bar{\xi}[t]\right)^2 \\ {\scriptstyle Eq.\ (30),\ 4} &= \frac{1}{2} \sum_{t=0}^{T} \Bigg( \gamma \mathbb{1}[t = 0] + \eta \mathbb{1}[t > 0] + \mu \bar{s}[t] \\ &\quad + C \sum_{\tau=1}^{t} \xi[t - \tau](\tau + c)^{-(1+\theta)} - \bar{\xi}[t] \Bigg)^2 \\ s.t.\ & \mu, \theta, C, c > 0 \end{aligned} \tag{34}$$

Note that Eq. (34) involves the model components $\xi[t - \tau]$ – as we will show in the Sec. 2.2, the objective function and its gradients are computed iteratively by estimating $\xi[\tau]$ from $\xi[1], \ldots, \xi[t-1]$. Also note that the recursive term is model estimates $\xi[t - \tau]$ rather than observations $\bar{\xi}[t - \tau]$, as we would like to have the model reproducing the whole observed time series, rather than predicting the next point given observed history. As will be discussed in Sec. 2.3, we further improve fitting stability by adding a $\mathcal{L}^2$ regularizer to the objective function.

## 2.2 Computing gradients

Eq. (34) is a non-convex objective, we use gradient-based optimization approach, and specifically `L-BFGS` [9] with pre-supplied gradient functions. We use the implementation supplied with the `NLopt` package [7] in R. We fit each video in parallel, starting with multiple random initializations to improve solution quality, and we present the solution with the lowest error function $J$. The gradient computations are listed as follows.

We define the error term as $e[t] = \xi[t] - \bar{\xi}[t]$, Eq. (34) now becomes $J = \frac{1}{2}\sum_{t=0}^{T} e^2[t]$. Since $\bar{\xi}[t]$ are observed quantities,

$$\frac{\partial e[t]}{\partial var} = \frac{\partial \xi[t]}{\partial var} \ ,$$

Here $var \in \{\mu, \theta, C, c, \gamma, \eta\}$. Using chain rule, we obtain:

$$\frac{\partial J}{\partial var} = \sum_{t=0}^{T} e[t] \frac{\partial \xi[t]}{\partial var} \tag{35}$$

Specifically, we compute the following partial derivatives and use them in Eq. (35) to compute the gradient.

$$\frac{\partial \xi[t]}{\partial \mu} = \begin{cases} \bar{s}[t] + C\sum_{\tau=1}^{t} \frac{\partial \xi[t-\tau]}{\partial \mu}(\tau + c)^{-(1+\theta)} & , t > 0 \\ \bar{s}[0] & , t = 0 \end{cases} \tag{36}$$

for $t > 0$,

$$\begin{aligned} \frac{\partial \xi[t]}{\partial \theta} &= C\sum_{\tau=1}^{t} \frac{\partial \xi[t-\tau]}{\partial \theta}(\tau + c)^{-(1+\theta)} \\ &\quad + \xi[t-\tau]\frac{\partial}{\partial \theta}(\tau + c)^{-(1+\theta)} \\ &= C\sum_{\tau=1}^{t} \left[ \frac{\partial \xi[t-\tau]}{\partial \theta} - \xi[t-\tau]\ln(\tau + c) \right] (\tau + c)^{-(1+\theta)} \end{aligned} \tag{37}$$

for $t = 0, \frac{\partial \xi[0]}{\partial \theta} = 0$ .

for $t > 0$,

$$\begin{aligned} \frac{\partial \xi[t]}{\partial C} &= \sum_{\tau=1}^{t} C \frac{\partial \xi[t-\tau]}{\partial C}(\tau + c)^{-(1+\theta)} \\ &\quad + \xi[t-\tau](\tau + c)^{-(1+\theta)} \end{aligned} \tag{38}$$

for $t = 0$, $\frac{\partial \xi[0]}{\partial C} = 0$ .

for $t > 0$,

$$\begin{aligned} \frac{\partial \xi[t]}{\partial c} &= C\sum_{\tau=1}^{t} \frac{\partial \xi[t-\tau]}{\partial c}(\tau + c)^{-(1+\theta)} \\ &\quad - (1+\theta)\xi[t-\tau](\tau + c)^{-(2+\theta)} \end{aligned} \tag{39}$$

for $t = 0, \frac{\partial \xi[0]}{\partial c} = 0$ .

For the unobserved external stimuli $\gamma$ and $\eta$.

for $t > 0$,

$$\frac{\partial \xi[t]}{\partial \gamma} = C\sum_{\tau=1}^{t} \frac{\partial \xi[t-\tau]}{\partial \gamma}(\tau + c)^{-(1+\theta)} \tag{40}$$

for $t = 0, \frac{\partial \xi[0]}{\partial \gamma} = 1$ .

for $t > 0$,

$$\frac{\partial \xi[t]}{\partial \eta} = 1 + C\sum_{\tau=1}^{t} \frac{\partial \xi[t-\tau]}{\partial \eta}(\tau + c)^{-(1+\theta)} \tag{41}$$

for $t = 0, \frac{\partial \xi[0]}{\partial \eta} = 0$ .

Note that the gradient computation is iterative, i.e. the computation of $\frac{\partial \xi[t]}{\partial var}$ makes use of previous values in its own series $\frac{\partial \xi[\tau]}{\partial var}$ for $\tau = 1, \ldots, t-1$.

## 2.3 Adding an $\mathcal{L}^2$ regularizer

We add $\mathcal{L}^2$ regularization on the linear coefficients of the Hawkes Intensity Process to avoid overfitting. The loss function with the regularization terms are as follows.

$$\begin{aligned} J_{reg}(\omega, \mu, \theta, C, c) &= J(\mu, \theta, C, c) + \\ &+ \frac{\omega}{2}\left( \left(\frac{\gamma}{\gamma_0}\right)^2 + \left(\frac{\eta}{\eta_0}\right)^2 + \left(\frac{\mu}{\mu_0}\right)^2 + \left(\frac{C}{C_0}\right)^2 \right) \ , \end{aligned} \tag{42}$$

Here $(\gamma_0, \eta_0, \mu_0, C_0)$ are reference values for parameters obtained by fitting the series $\bar{\xi}[t]$ without regularization. The reference values are used to normalize the parameters in the regularization process, so that they have equal weights. Intuitively using $\mathcal{L}^2$ normalization in square-loss is effectively putting a Gaussian prior on the parameters being regularized. We desire parameters $c$ and $\theta$ to take values away from zero, hence they are not regularized. The $\mathcal{L}^2$ regularization term is differentiable with respect with variables $(\gamma, \eta, \mu, C)$ and the terms $\frac{\omega}{\gamma_0}$, $\frac{\omega}{\eta_0}$, $\frac{\omega}{\mu_0}$ and $\frac{\omega}{C_0}$ are added respectively to the RHS of Eq. (40), (41), (36) and (38).

The regularizer parameters $\omega$ is expressed as a percentage of $J_0$ (the value of the non-normalized error function) and it is determined through a line search within $[10^{-4}J_0, 10J_0]$ in log-scale. $\omega$ is tuned per video, on a temporally hold-out tuning sequence, i.e. we use the first 75 days of observed popularity for parameter estimation, the next 15 days for tuning $\omega$, and day 91-120 for forecasting popularity.

### 2.4 Properties of the model estimates

It is informative to discuss the properties of the model estimation procedure above in terms of model properties, and the optimization procedure.

The Hawkes intensity model in Eq. (30) is a non-linear integral equation. It is worth noting that there are two non-linear parameters $\theta$ and $c$, and the rest are linear parameters – $\mu$ and $C$, as well as the unknown external stimuli $\eta$ and $\gamma$. Given $\theta$ and $c$, the loss function in Eq. (34) is convex, and the optimization procedure converges to a global optima. Assuming a set of fixed (or known) non-linear parameters for the whole dataset is therefore convenient for fast estimation, and is used in recently literature such as by Zhao *et al.* [16]. On the other hand, our own recent study [11] shows that in addition to better interpretability, there is a performance advantage of estimating both the linear and non-linear parameters in estimating Hawkes point processes. Therefore we estimate all of the linear and non-linear parameters for the Hawkes intensity process.

The procedure for minimizing the squared loss in Eq. (30) uses a standard gradient-based non-linear continuous optimization routine. The procedure will converge to a local minima in the loss function when it terminates. We implement standard random restarts to improve the solution quality, i.e. perform the optimization from 8 different random starting points for each video, choose the one with the lowest loss as the final result. On the theoretical end, there is no known results of convergence rates as a function of sequence length (or sample size) for this class of models. In practice, the primary limiting assumptions is the model being stationary (and fixed) over time and over different parts of the activated online social network.

## 3 Data

**Setting up the Twitter crawler** We construct a "Tweeted Videos" dataset using the data APIs from both Twitter and YouTube. We stream tweets from Twitter API using the set of keywords related to YouTube and its video: `"youtube" OR ("youtu" AND "be")`. The Twitter filter API returns the tweets for which the keywords were matched in at least one of the considered fields, including in the textual description and the `expanded_url` field. Twitter API claims that the `expanded_url` field contains the original URL of all URLs shortened using shortening services (such `bit.ly`). While this happens in most of the cases, we found that a non-trivial number of URLs remain shortened. In these cases, the Youtube video ID is hidden, if the URL is a link towards a Youtube video at all. Expanding these links ourselves is unfeasible, given our network and service constraints. One noteworthy exception is Youtube's own shortening service (`youtu.be`) which readily contains the video ID. It is for this reason that we added `"youtu" AND "be"` to the filter keywords.

This returns over 5 million matched tweets per day after URL expansion and tokenization performed by Twitter, most of which mention and link to a YouTube video. The raw dataset used in this study was from 2014-05-29 to 2014-12-26, having 1,061,661,379 tweets in total. From each of these tweets, we extracted the associated YouTube video id (only the first in case multiple videos were referenced in the same tweet), resulting in 81,915,174 distinct videos in total.

**Setting up the Youtube crawler** From YouTube.com, we obtained for each video its metadata, including the upload date and video category, as well as the time series consisting of the daily number of views and shares, from the day of upload and until the date of crawling. We aggregate for each video the number of tweets it receives every day and we obtained three attention-related time series for each video: ($views[t]$, $shares[t]$ and $tweets[t]$), here $t$ indexes time with unit of a day. For the number of views and shares, the time range is from the video's upload date to the data collection (i.e. February-March 2015). For number of tweets, time ranges from the videos upload date or 2014-05-29 (whichever is later) to 2014-12-26.

### 3.1 The 5Mo and Active datasets

We constructed two cleaned data subsets from the feed of tweeted videos, in order to collect basic data statistics and estimate the model.

- The 5Mo was constructed to have videos whose popularity history is at least 60 days long, and is used for forecasting popularity. We narrow down the timeframe of video upload to between 2014-05-29 and 2014-10-24 in order to have long enough history. There are 16,417,622 videos with publicly-available popularity history. We did not obtain the popularity history for more than half of the videos, reasons for such data loss include: a video is no longer online, a video's popularity history is not publicly-available, or requests that resulted in web server errors. This large and diverse sample allows us to estimate the background statistics of video views, tweets, and shares, as will be discussed in the next subsection.
- The Active dataset selects videos uploaded between 2014-05-29 and 2014-08-09, and which have received

Table 1: Number of videos broken down by category in the ACTIVE dataset. `Music` represents a significant proportion (25%) of all the videos.

| Category | #vids | Category | #vids |
|---|---|---|---|
| Comedy | 865 | Music | 3549 |
| Education | 298 | News & Politics | 1722 |
| Entertainment | 2422 | Nonprofits & Activism | 333 |
| Film & Animation | 664 | People & Blogs | 1947 |
| Gaming | 882 | Science & Technology | 262 |
| Howto & Style | 180 | Sports | 614 |
| Total: | 13,738 | | |

at least 100 tweets and at least 100 shares during recorded lifetime. The timeframe is selected to ensure that each video has at least 120 days of tweeting and popularity history in our dataset, the activity threshold is used to ensure there is sufficient data for estimating the Hawkes intensity model and producing a forecast as describe in Section 5. Table 1 presents the category distribution for the ACTIVE dataset. It is noteworthy that the largest 4 categories cover more than 70% of all the videos in ACTIVE, with more than 25% of the videos being `Music`. We removed 6 content categories (i.e. `Autos & Vehicles`, `Travel & Events`, `Pets & Animals`, `Shows`, `Movies`, `Trailers`) containing less than 1% of the videos in the dataset. Their corresponding videos were also removed. The resulting dataset contains 13,738 videos.

## 3.2 The popularity scale over time

It is well-known that network measurements such as the number of views follows a long-tailed distribution. To facilitate discussions about popularity and attention, we propose to quantify popularity on an explicit percentage scale, with 0.0% being the least popular, and 100% being the most popular. Videos are grouped into *Popularity Bins* by viewcounts that they receive at $t$ days after upload, and each bin $\xi_t(k)$ is marked with its maximum popularity percentile – videos in bin $k$ are at most among the top $k$% popular with age $t$. In this work we use 40 evenly spaced bins, *i.e.*, $k = 2.5, 5.0, 7.5, \ldots, 100$, and each bin contains 2.5%, or ~41K+ videos for 5MO.

Fig. 8(a) and (b) contain boxplot of video viewcounts (in log-scale) of each bin after 30 and 60 days, respectively. We can see the long tailed distribution of popularity in YouTube reflected here – videos in the less popular bins have very similar number of views, *e.g.* the first 6 bins, or 15% of the videos, all have less than 10 views;

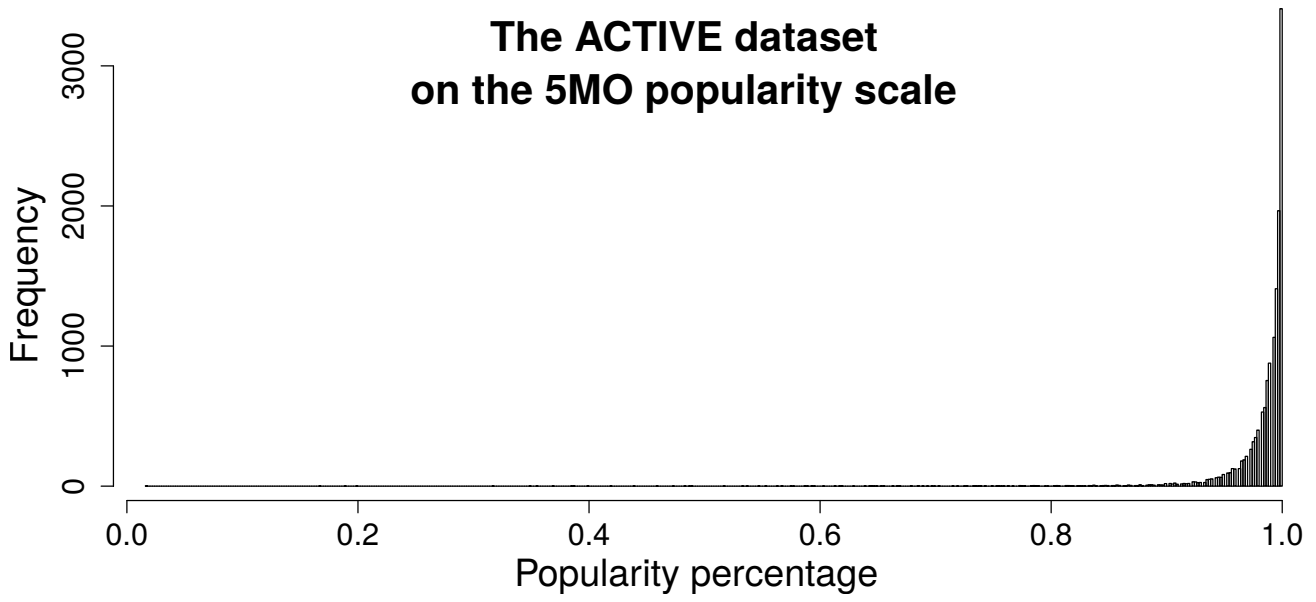

Figure 7: Positioning of the ACTIVE dataset on the popularity scale of the 5MO dataset (at 30 days after upload). The horizontal axis shows the popularity percentiles in the 5MO dataset, while the vertical axis shows the corresponding frequency of videos in ACTIVE. Visibly, ACTIVE is a subset of the most popular videos in 5MO.

videos in each the middle bins (*e.g.* $k = 17.5, \ldots, 85.0$) are within 1.5 times of the view count of each other; yet viewcounts of the 5% most popular ($k = 97.5$ and $k = 100$) videos span over almost two orders of magnitude. For videos in 5MO at 30 and 60 days after upload, the shape of the overall popularity scale remains the same, with a slight increase in the dynamic range of views (top of the last boxplot). The popularity scale of the ACTIVE is very similar to the one presented in Fig. 8(a) and (b), the only notable difference being the number of views corresponding to each bin. ACTIVE is a subset of the most popular videos, as shown by Fig. 7: the videos in ACTIVE are positioned in the top 5% popularity percentiles of 5MO ($k = 97.5$ and $k = 100$).

In Fig. 8(c) we explore the change of popularity of each video from 30 days (y-axis) to 60 days (x-axis). Note that most videos retain a similar rank (in the boxes along the 45 degree diagonal line), or have a slight rank decrease as they are overtaken by other videos (slightly above the diagonal in the plot). No outliers exist in the upper-left part of the graph, since a video cannot lose viewcount that it already gained. Most notably, we can see that video from any bucket can *jump* to the top popularity buckets between 30 and 60 days of age, such as the outliers for the few boxes on the far right. This phenomenon elicits important questions: how did these videos do viral, and whether or not it is related to external promotions.

# 4 Understanding popularity dynamics

In this section, we provide additional observations on the parameters the HIP model, supplementing the analysis presented in the main text Sec. 4. Specifically, we relate the distribution of specific parameter values such as memory exponent or exogenous sensitivity, to video groups – channels, content categories – and a video's popularity.

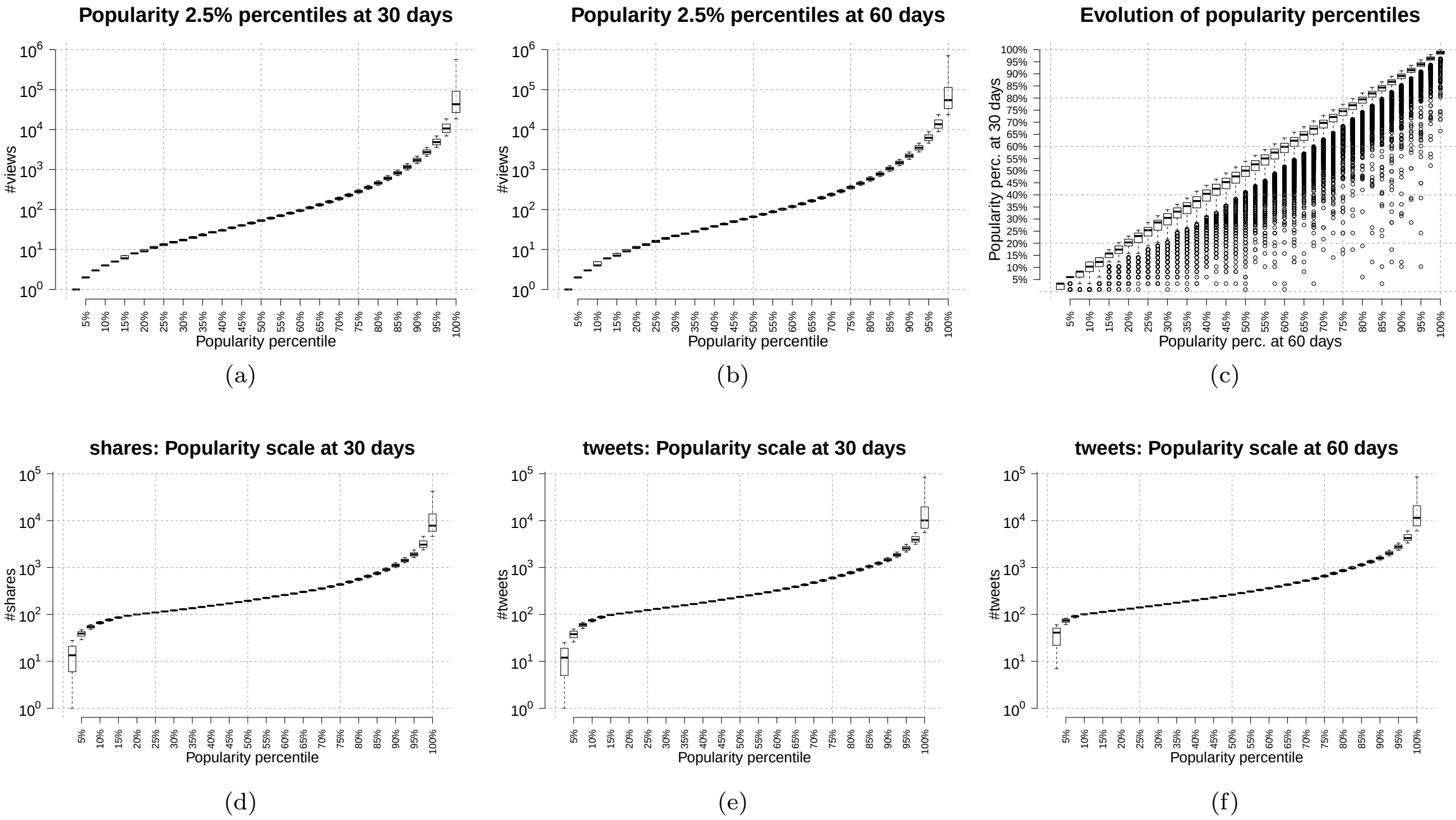

Figure 8: **(top row)** The popularity scale of YouTube videos in the 5Mo dataset. The total number of `views` obtained by each video in the first 30 days (a) and 60 days (b) after upload is divided into 40 equally spaced bins (i.e. each with 2.5% of the videos). The 2.5% most popular videos span almost two orders of magnitude in views. Note that outliers in this bin are not represented, as the most popular videos in the collection have $\sim 10^8$ views. (c) Evolution of popularity between 30 and 60 days. The outliers are videos that have improved significantly on the popularity scale. **(bottom row)** The popularity scale on the Active dataset for `shares` at 30 days (d) and `tweets` at 30 days (e) and 60 days (f). Note that the scales for `shares` and `tweets` are very similar, with the observation that videos in the Active set seem to receive more tweets than shares. Another observation is the difference in the popularity scale for tweets between 30 and 60 days: the biggest change is observe in the bottom 2.5%. The reader should remember that all videos in the Active set receive at least 100 tweets during their life time. As a result, the bottom 2.5% will continue to rise with $t$.

## 4.1 Behavior across groups of videos: categories and channels

We provide in this subsection some observations on behavior statistics and key parameters broken down by video category. Furthermore, we show how the endo-exo map can be used to detect consistent behaviors across YouTube channels.

**Consistent behavior across channels** We use the endo-exo map to visualize groups of videos that belong to the same user-assigned content type, or are from the same author, called *channel* in YouTube. Fig. 9 shows a scatter plot of videos posted by a reporter in category `News & Activism` (in red) and a user focusing on recordings of `Game` sessions (in blue). The game recording videos are generally more popular (bigger circles) than the news videos, and this is explained by the former group having higher exogenous sensitivity – higher values of $\mu$.

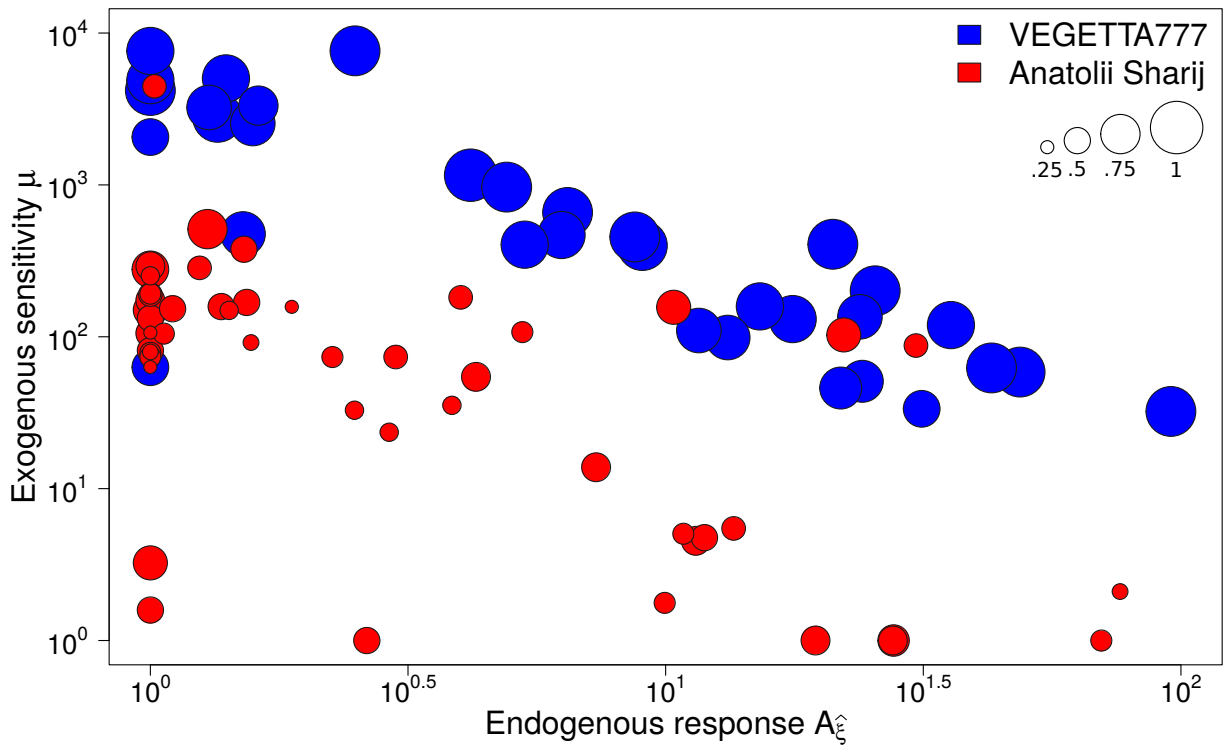

Figure 9: Video channels on the endo-exo map. Scatter plot of videos from a reporter covering events in Ukraine (Anatolii Sharij, in red) and Spanish game recording videos channel (VEGETTA777, in blue). Radii of the circles are proportional with the popularity percentile of each video.

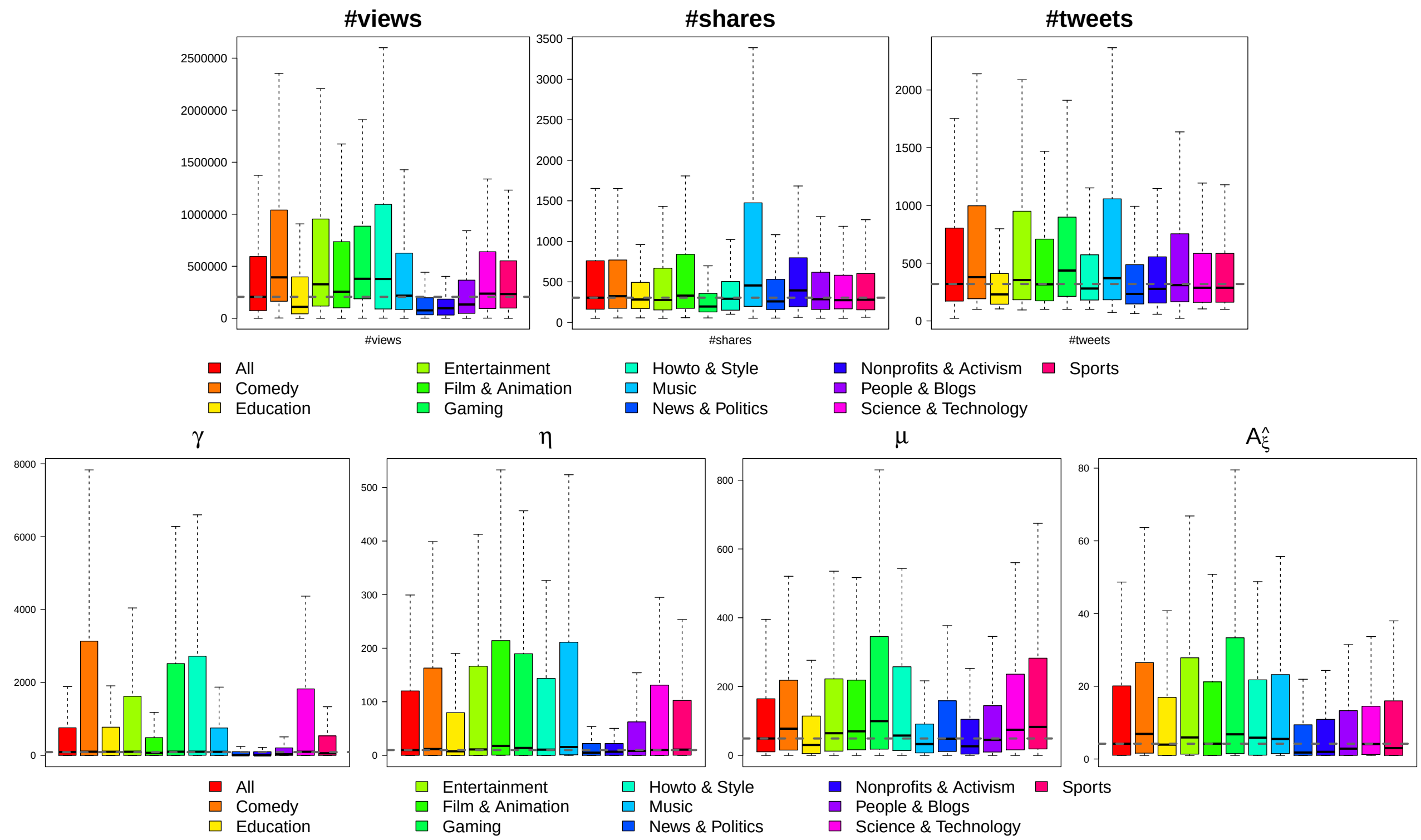

Figure 10: (Top row) The number of views (left), shares (center) and tweets (right) for videos in different categories. (Bottom row) Box plots of unobserved exogenous influence (initial impulse $\gamma$, constant excitation $\eta$), exogenous sensitivity $\mu$ and endogenous response $A_{\hat{\xi}}$, broken down by category.

**The effect of the external influence is not equal** We examine the amount of attention (in number of views) and external influence (in number of shares and tweets) in the ACTIVE dataset. This provides a basis for understanding the corresponding Hawkes intensity model. Fig. 10 (top row) contains box plots of total views, along with total shares and tweets, broken down by video category. The left-most boxes (in red) depicts the profile of all videos. One notable example is videos in the `Nonprofits & Activism` category: overall they have less-than-average amount of views, despite being shared more than the median number of times.

**Observed versus unobserved external influences** Model parameters $\gamma$ and $\eta$ can be interpreted respectively as the initial impulse and constant exogenous stimuli not captured in the observed exogenous activity $s(t)$. From the bottom left two plots in Fig. 10, we can see that several categories have significantly higher components of $\gamma$ and $\eta$, such as `Gaming`, `Comedy` and `Entertainment`. This may result from a significant volume of activity outside of Twitter or Youtube sharing – `Gaming` videos, for example, is known to spread on dedicated social networks such as sub-reedit `/r/gaming/`, `/r/gamingvids/` or forums, such as www.minecraftforum.net.

**Exogenous sensitivity and endogenous response** The two bottom right plots of Fig. 10 represent the breakdown per category of, respectively, the exogenous sensitivity $\mu$ and the endogenous response $A_{\hat{\xi}}$. These plots present an alternative view to the 2-dimensional density distribution of each category on the *endo-exo map*, shown in Figures 16 and 17. Certain categories, such as `Comedy`, `Gaming` or `Sport` seem to be particularly sensitive to external influence. Categories like `Comedy`, `Entertainment` or `Gaming` observe higher then median endogenous responses. The fact the `Comedy` and `Gaming` show both a high exogenous sensitivity and endogenous response provide a plausible explanation to why these categories observe relative high popularity (`#views`) despite their relative low sharing. Conversely, `Nonprofits & Activism` exhibits lower than median values for both $\mu$ and $A_{\hat{\xi}}$ which accounts for its low popularity (even though highly shared).

## 4.2 Categories of longer versus shorter memory

We study the distribution of the memory exponent $\theta$ in the HIP model, for three categories of videos in the ACTIVE dataset. In Fig. 11, the distributions for categories `Music`, `Nonprofits & Activism` and `News & Politics` (in red) are contrasted with the distribution from `All` the videos (in blue). The solid lines in each graph indicate the median value for $\theta$ in each category, whereas the dashed lines indicate the mean value. All video categories, as well as the general population, observe a long tail distribution for $\theta$, with a peak density around $\theta \simeq 3.35$. A small $\theta$ leads to slower decay over time (and larger endogenous response $A_{\hat{\lambda}}$), whereas a large $\theta$ means an video is quicker to be *forgotten* (i.e. small $A_{\hat{\lambda}}$). We can see that a larger (than random) fraction of `Music` videos decay slowly ($\text{mean}_{\theta,all} = 15.94$, $\text{mean}_{\theta,music} = 14.95$), while more `News & Politics` and `Nonprofits & Activism` videos are forgotten faster, with $\text{mean}_{\theta,nonprofit} = 17.56$ and $\text{mean}_{\theta,news} = 19.45$. This suggests that there is a systematic difference across different types of videos in the rate at which the collective *memory* decays – one explanation for such differences can be that music is typically considered *timeless* content while news is considered *timely* whose relevance decreases rapidly over the first few days.

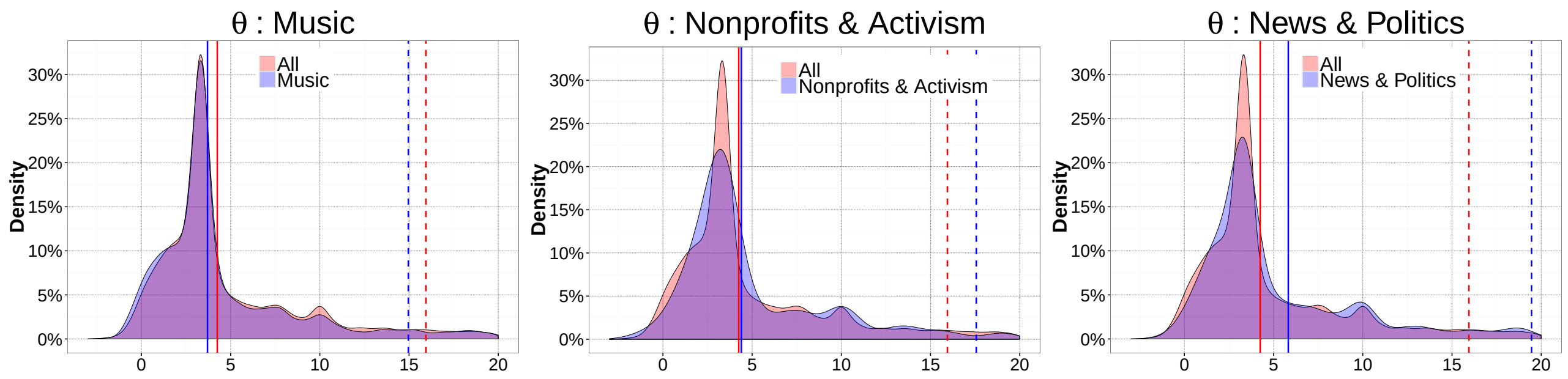

Figure 11: Distribution of the memory exponent $\theta$ for 3 categories: `Music` (a), `Nonprofits & Activism` (b) and `News & Politics` (c), compared to the background distribution in all videos. Solid vertical lines indicate the median $\theta$ value in each population, whereas the dashed vertical lines indicate the mean $\theta$. Color of lines corresponds to legend. The densities are obtained using kernel density estimation.

## 4.3 Model parameters and popularity

In this section, we provide additional details about the relation between video popularity and fitted values of parameters $\mu$ and $\theta$. These analyses provide additional details to the endo-exo map, by explicitly linking the endogenous and exogenous components of each video to each model parameter.

**Parameters $\mu$ and $\theta$ and popularity** In the main text, we claim a direct connection between $\mu$ the exogenous sensitivity and popularity and an inverse connection between the $\theta$ the time-decay rate of the memory kernel and the popularity. We provide, in Fig. 12, empirical proof of these connections by studying the popularity distribution for low and high values of the above parameters. The top-left graphic shows the density distribution of the fitted values of $\mu$ in the Active dataset. There is a high a peak of density around $\mu = 1$, corresponding to videos with low sensitivity to external influence, and a second peak around $\mu = 10^{1.73} = 53.7$, corresponding to videos with higher exogenous sensitivity. We divide the range of $\mu$ into deciles (groups of 10% each) and we select the second decile (i.e. low sensitivity) and the tenth decile (high sensitivity), hashed in gray on the graphic. In the bottom-left graphic, we plot the popularity distribution for videos within each of the above deciles of $\mu$. The subpopulation of videos with low exogenous sensitivity show a dense area of low popularity, and with only very few videos making it into the top popularity percentiles. Conversely, the density distribution of the subpopulation of videos with high exogenous sensitivity shows an increasing trend, with a concentration of highly popular videos. This confirms the intuition that highly popular videos tend to have high values of exogenous sensitivity $\mu$.

Similar results can be shown for the time-decaying memory exponent $\theta$, which controls how fast videos are *forgotten* and the size of the endogenous response $A_{\hat{\lambda}}$. Fig. 12b plots the density distribution of $\theta$, which shows a peak at $\theta = 3.36$ and selects the second and tenth percentile, corresponding respectively to low values and high values of $\theta$. Similarly to $\mu$, the bottom-center graphic plots the popularity distribution for each of the subpopulations defined by the selected deciles of $\theta$. The subpopulation with high values of $\theta$ (i.e. low $A_{\hat{\lambda}}$) tends to be forgotten more quickly and shows a concentration of videos with low popularity, whereas videos with lower values of $\theta$ (and higher $A_{\hat{\lambda}}$) tend to be more popular.

**Endo-exo map for additional categories** The above considerations are at the basis of the construction of the *endo-exo map*, as shown in the main text and its potentially viral region – videos with high values of both exogenous sensitivity $\mu$ and endogenous response $A_{\hat{\lambda}}$ are more susceptible to become popular *if given the required attention*. The right column of Fig. 12 plots the 2D density of videos on the endo-exo map for the entire Active dataset (top) and the top 5% most popular videos (the color map is aligned for the two graphics). Visibly, the distribution of the popular videos is skewed towards the more viral region of the map (i.e. high $\mu$ and high $A_{\hat{\lambda}}$). In Fig. 16 and 17, we repeat this analysis and we further break down the Active population, based on video category. We plot pairs of 2D densities of videos on the endo-exo map for all categories, except `Gaming` and `Film & Animation`, which were discussed in the main text Fig. 4. This visually reveals some of the dynamics that propel videos to the most popular segment, in each subpopulation. For example, categories like `Gaming`, `Science & Technology` `Travel & Events` have the distribution of the most popular videos shifted top-right w.r.t to rest of the category (similar to the dynamics shown for the entire population). Other categories appear only upward-shifted (i.e. only higher $\mu$) w.r.t to rest of the category: `Film & Animation`, `Entertainment`, `Howto & Style`, `News & Politics` and `People & Blogs`. There is even an outlier category, `Comedy`, which seems to have two heat centers in the top 5% popular subpopulation. This seems to indicate two distinct patterns of becoming popular within this category: one pattern involves being sensitive to exogenous excitation more than the average video, whereas the second pattern involves higher endogenous propagation in the network (higher $A_{\hat{\lambda}}$).

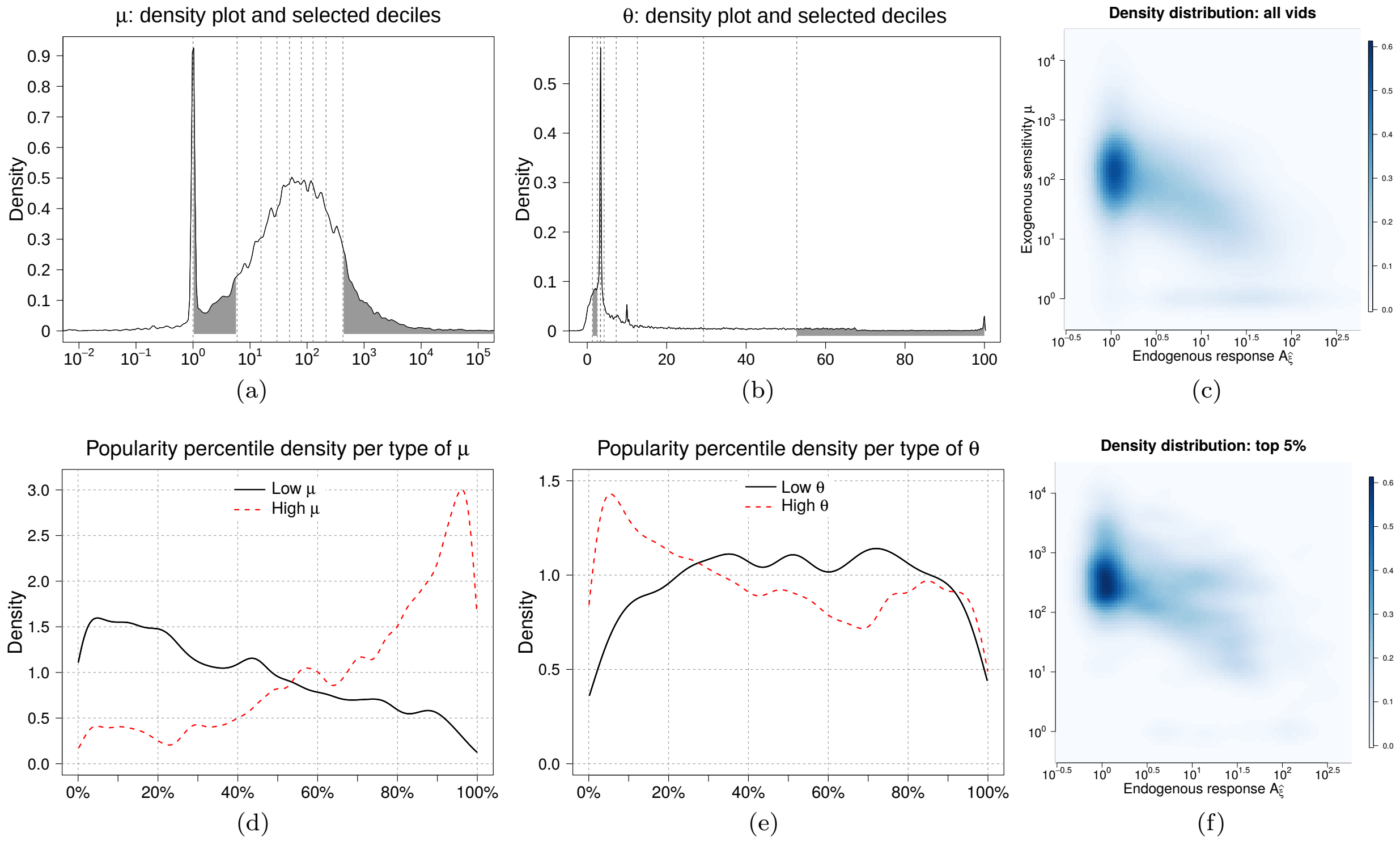

Figure 12: Density distribution of fitted model parameters values $\mu$ (left column) and $\theta$ (center column). The range of each fitted parameter is divided into 10 deciles, shown by vertical dotted gray lines (in the top row). For each of parameters $\mu$ and $\theta$, 2 deciles are chosen (one corresponding to high values, and a second one corresponding to low values), shown shaded in gray. For each parameter, the bottom row plots the popularity distribution for videos within each of the chosen deciles. Right column: the *endo-exo map* 2 dimensional density distribution for all videos (top) and top 5% most popular videos (bottom). The density distribution of most popular videos is skewed towards the more viral area of the map (high $\mu$ and high $A_{\hat{\xi}}$).

## 4.4 Potential causal connection between the views, tweets and shares series

In this section, we investigate the causal connection between the series of views, shares and tweets, for each of the videos in the Active dataset. We test for causality using time-series analysis tools. We employ a F-type Granger-causality test [4], implemented in the R package `vars` [13]. For each series of each video we construct a Vector Autoregressive Model, with the lag determined automatically using minimal AIC. Next, we perform a Granger causality test for each video and each pair of temporal series – i.e. (views, shares), (views, tweets) and (tweets, shares). Each test is performed in both directions – e.g. views Granger-cause shares and shares Granger-cause views. The null hypothesis is that no causal relation exists between the series. We reject the null hypothesis and we accept the existence of a causal relation when we observe a test p-value lower than $10^{-3}$.

Table 2: Granger causality test: number of videos in the Active set for which the null hypothesis (i.e. absence of Granger causality) is rejected with $p < .001$. The left side of the table shows the number of videos for which an unidirectional Granger causality is detected – i.e. for 114 videos tweets Granger-cause shares and not the other way around. The right side shows the number of videos for which the relation is detected in both directions. Note that when a video presents a bidirectional relation, it is not counted the unidirectional relations for the same pair of series.

| | unidirectional | | | bidirectional | | |
|---|---|---|---|---|---|---|
| | tweets | shares | views | tweets | shares | views |
| tweets | - | 114 | 164 | - | 22 | 18 |
| shares | 136 | - | 537 | | - | 253 |
| views | 162 | 833 | - | | | - |

Tab. 2 shows the number of pairs in each setup for which the causal relation is considered to be significant. Note that the causal relation can be reciprocal – e.g. for 253 videos, both the shares Granger-cause the views and and the views Granger-cause the shares. Considering the scale of the Active dataset (around 14 thousands videos), a causal relation is detected for no more than 6% of the videos – i.e. for the relation views Granger-cause shares, true for 833 videos. For all pairs of series, the number of videos presenting a unidirectional relation seems comparable (e.g. tweets Granger-cause views for 164 videos, and shares Granger-cause tweets for 162 videos). We cannot identify a clear Granger causality relation between different series. As there does not yet exist a standard method for capturing non-linear causal relationships or causality with confounding effects, we leave this as future work.

# 5 Popularity forecasting and comparison to baseline

In this section we provide additional details and results to complete the analysis in the main text Sec. 5. Namely, we provide more information about the performance break down of different approaches and the statistical testing analysis for detecting statistically significant differences in the forecasting performance.

The series of the first 90 days of each video history in Active dataset are used to fit the Hawkes intensity model parameters. The series is divided into two sub-series: the first 75 days are used to fit parameters $\{\mu, \theta, C, c\}$, while the last 15 days for the holdout series used to fit the regularizer meta-parameter $\omega$. Either `#shares` and `#tweets` series can serve as the known exogenous stimuli series $s(t)$. The Multi-Linear Regression (MLR) [14] baseline is trained using the same data. We adapt the original algorithm by predicting the value of the viewcounts for each of the 30 days between day 91 and 120. Furthermore, we build an enhanced version (denoted by *MLR (#shares)* or *MLR (#tweets)*) by introducing the exogenous influence as additional variables, both in the training and in the prediction. The baseline is particularly sensitive to outliers, which we remove from the training set. A video is considered an outlier if it has received a large burst of views in the period from 91 to 120 days. More precisely, we remove any video having received twice as many view between days 91 and 120 than then do between 61 and 90. 3.5% of the videos are considered outliers and eliminated from the training set. The errors are measured in average error in popularity percentile, as defined in the main text.

## 5.1 Additional results

In addition to the performance comparison, shown in the main text Fig. 5, Fig. 13(a) presents the Cumulative Distribution Function (CDF) of the prediction errors for the HIP, MLR and MLR (`#shares`). HIP consistently outperforms MLR (with and without the exogenous stimuli information): HIP forecasts popularity of 87% of the video population with a maximum 10% error, while MLR covers only 78% of the population for the same error threshold. Furthermore, MLR (`#shares`) obtains only marginal performance improvements over MLR, even while using the exogenous information. Fig. 13(b) shows the absolute forecasting error performances, aggregated using barplots. Visibly, the HIP (using either `#shares` and `#tweets`) consistently outperforms MLR both in term of median values and variation, which results in the better mean values of forecasting error already shown in the main text Fig. 5(center). Fig. 13(c) analyzes closely the forecasting error distribution for the best performing version of each approach. The HIP (#shares) blue curve shows a concentration of lower errors and a median error value of 3%. The red curve corresponds to the error distribution for the MLR (#shares) and shows a higher concentration of larger error and a median error value of 3.75%.

## 5.2 Comparing performance

We ask whether or not the performance differences observed in main text Fig. 5(b) are significant, or are they due to chance. We break down the study into two questions: 1) is the difference of forecasting performance between the Hawkes intensity process and the MLR baseline statistically significant? and 2) does the source of exogenous stimuli – `#shares` or `#tweets` – influence the quality of the forecasting? We setup four comparisons: two comparing the performances of forecasting methods for each of the two sources of influence and two comparing the effect of the sources for each of the two algorithms. Based on the selected setup, we construct two samples and perform a T-test. For each test, the null hypothesis assumes that the mean of two samples are equal (i.e. there is no difference in forecasting performance). The alternative hypothesis assumes that the true means of two samples are not equal. Note that these statistical tests are not connected to particular models that were used to obtain the estimates, i.e. it does not matter if the estimated come from the multivariate linear regression or from the Hawkes intensity process.

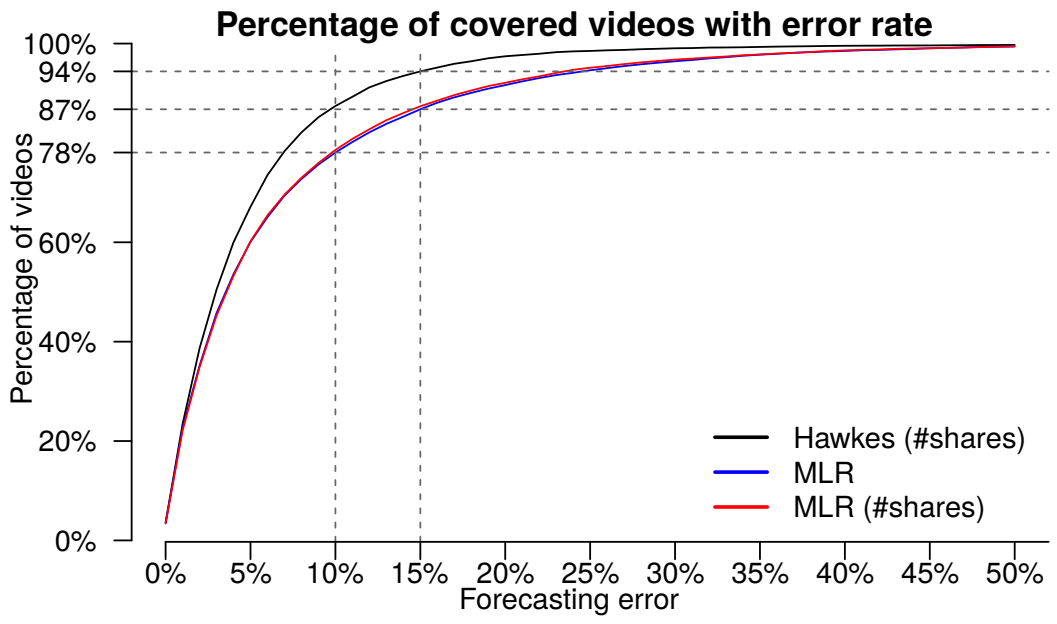

(a)

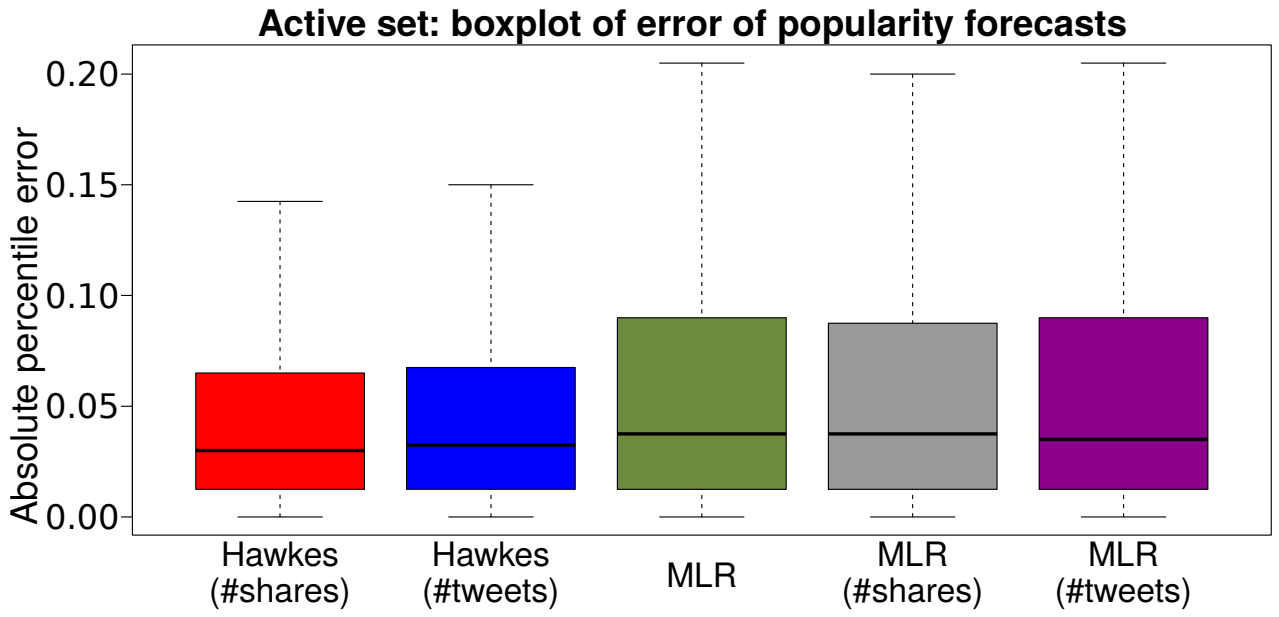

(b)

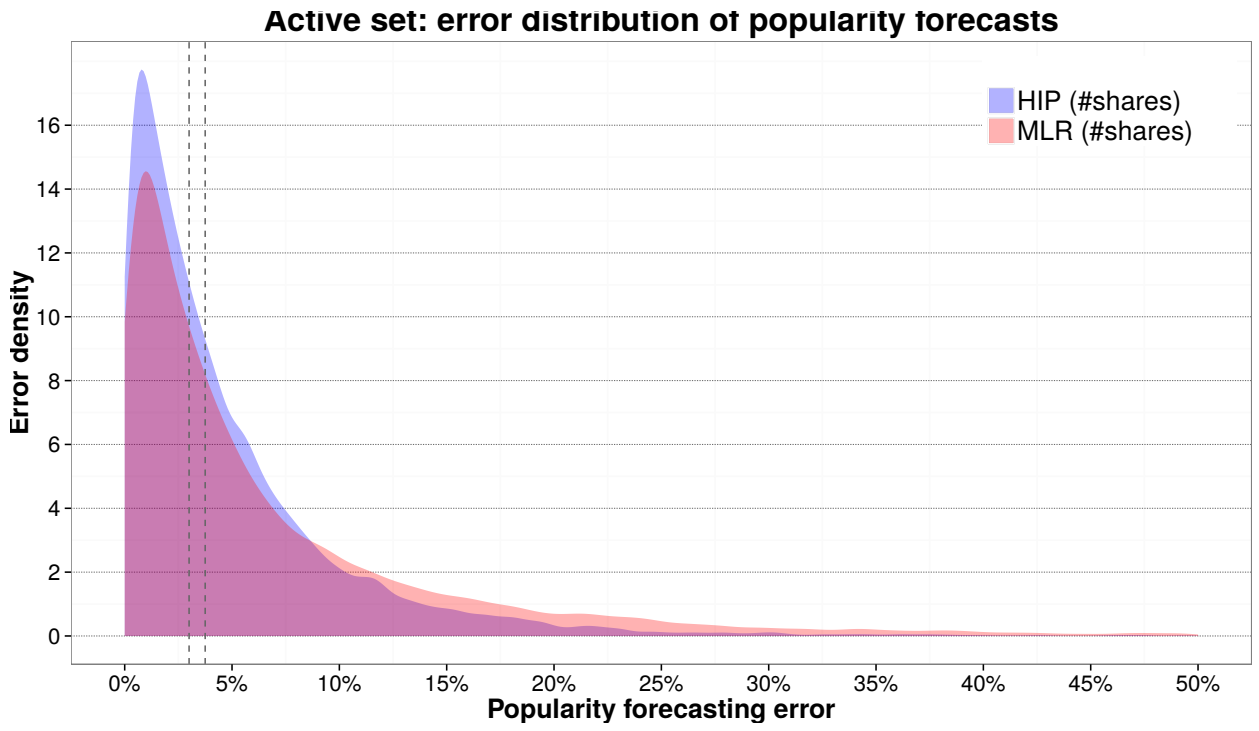

(c)

Figure 13: Performance comparison graphics, additional to main text: number of covered videos, when accepting a given error percentage (left) and barplots of absolute forecasting error (right). (bottom) Absolute forecasting error distribution for Hawkes (`#shares`) (blue curve) and MLR (`#shares`) (red curve). Median values are represented for each approach with gray vertical lines: 3% for Hawkes and 3.75% for MLR.

**Statistical testing with large sample size** The well-known *p-value* issued from hypothesis testing is dependent on the observed difference between the two samples, as well as the sample size. This renders analyses based only on p-value particularly sensitive to sample size, given that with sufficiently large samples, a statistical test will almost always demonstrate a significant difference [15]. Given the size of the ACTIVE dataset (i.e. 13, 738) which serves as sample for the four experiments hereafter, we measure the *effect size* in addition to the typical p-value. The effect size measure which we report and we use to justify our analysis is *Cohen's d coefficient* [2], defined as the difference of the means scaled by the inverse of the standard deviation of both populations

**Evaluation setup** Our forecasting systems uses two inputs: the observed `#views` and an external stimuli source (either `#shares` or `#tweets`). Answering question 1) – significance of performance difference between the Hawkes intensity model and the MLR baseline – boils down to comparing two treatments for a single set of individuals. This translates into applying a *paired T-test* to a single sample. Conversely, comparing the two sources of exogenous excitation involves applying the same forecasting method to two different populations. This leads to applying a *two-sample T-test*.

**No performance difference between the two exogenous stimuli sources** The detailed results of each of the four hypothesis testings are presented in Tab 3. The first two lines correspond to the two-sample T-tests, dealing with the difference between the two sources of exogenous stimuli. The first test uses HIP as forecasting algorithm, the second one uses the MLR baseline. For both tests, the Cohen's d coefficient has a negligible value. This suggests that no significant difference exists when forecasting future popularity using `#shares` or `#tweets` as external stimuli. The very small p-value yielded in the first test ($\sim 10^{-7}$) is most likely an artifact of the large sample size. These two results indicate that the same forecasting performance is achieved using either source of external stimuli, regardless of the forecasting algorithm. Furthermore, a correlation analysis between the two sources reveals the same conclusion: for each video we compute the Pearson's correlation coefficient between the two time series `#shares` and `#tweets`. Fig. 7 shows the density distribution of the correlation coefficient. Visibly, the two series are highly correlated for most videos.

Table 3: Summary of comparison for forecasting performance differences. Each line corresponds to a performed T-test, either comparing two exogenous sources or two forecasting algorithms. Columns "Sample **A**" and "Sample **B**" describe the two compared samples in terms of used algorithm and exogenous source, mean value and standard deviation. The first two tests are *two sample T-tests*, whereas the last two lines correspond to *paired T-tests* (more details about the underlying rationale in the text). $M$ denotes sample mean, $SD$ standard deviation.

| Exogenous source | Forecasting algorithm | Sample **A** | Sample **B** | T-test p-val | Cohen's $d$ |
|---|---|---|---|---|---|
| `#shares` *vs.* `#tweets` | HIP | HIP (`#shares`) $M = 4.96 \times 10^{-2}, SD = 6.3 \times 10^{-2}$ | HIP (`#tweets`) $M = 5.35 \times 10^{-2}, SD = 6.88 \times 10^{-2}$ | $6.83 \times 10^{-7}$ | $-0.05$ |
| `#shares` *vs.* `#tweets` | MLR | MLR (`#shares`) $M = 6.94 \times 10^{-2}, SD = 9.08 \times 10^{-2}$ | MLR (`#tweets`) $M = 6.94 \times 10^{-2}, SD = 9.05 \times 10^{-2}$ | 0.98 | 0.00 |
| `#shares` | HIP *vs.* MLR | HIP (`#shares`) $M = 4.96 \times 10^{-2}, SD = 6.3 \times 10^{-2}$ | MLR (`#shares`) $M = 6.94 \times 10^{-2}, SD = 9.08 \times 10^{-2}$ | $8.57 \times 10^{-151}$ | 0.253 |
| `#tweets` | HIP *vs.* MLR | HIP (`#tweets`) $M = 5.35 \times 10^{-2}, SD = 6.88 \times 10^{-2}$ | MLR (`#tweets`) $M = 6.94 \times 10^{-2}, SD = 9.05 \times 10^{-2}$ | $2.41 \times 10^{-95}$ | 0.197 |

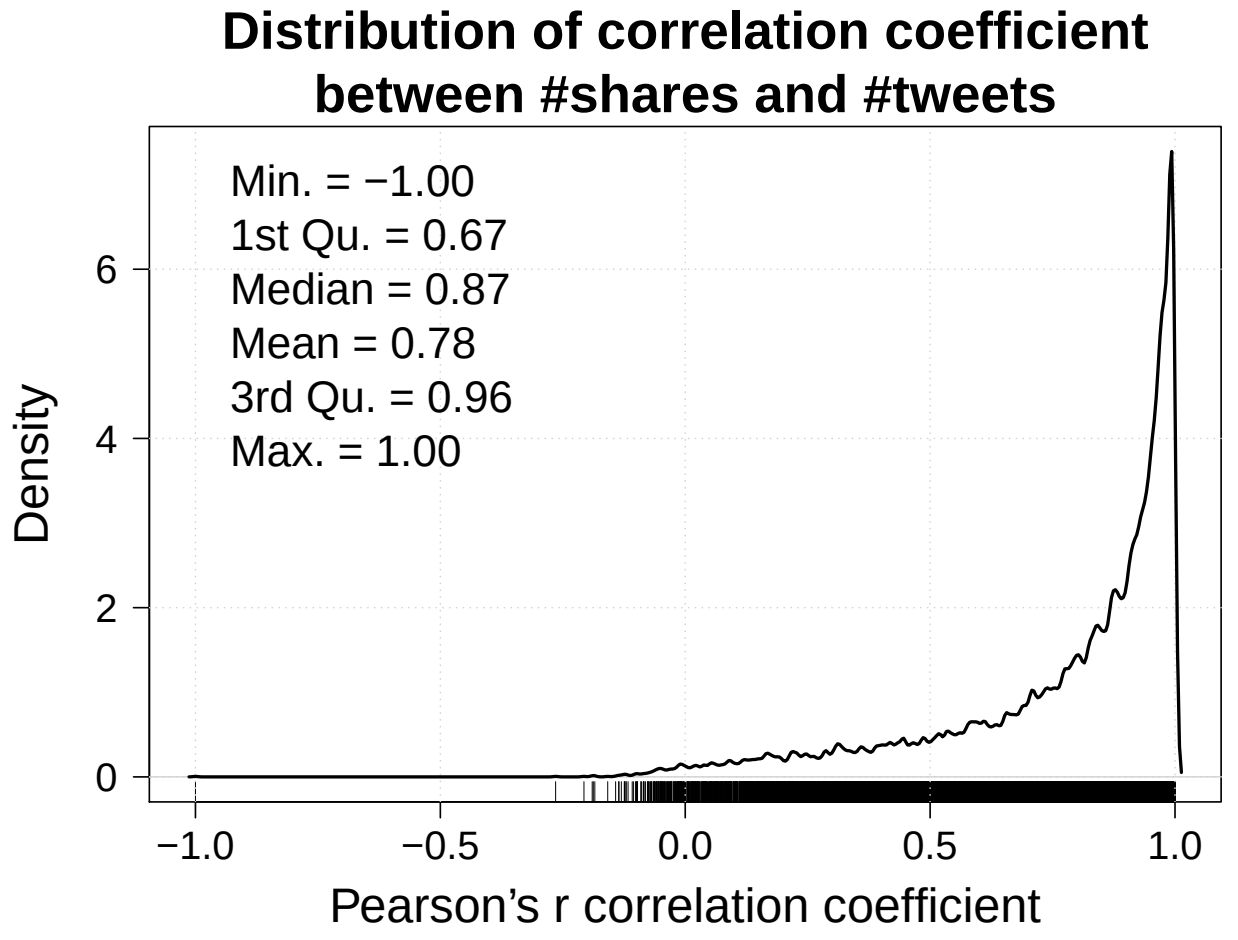

Figure 14: Density distribution of Pearson's $r$ correlation coefficient between `#shares` and `#tweets` for each video in the Active dataset. The legend values give the quartiles and the mean of the obtained coefficients.

**Differences between HIP and MLR** The last two lines of Tab. 3 show the results of testing the performances of the HIP model against the baseline, for each of the two external sources. In both cases, the Cohen's d coefficient shows a small, but non-negligible effect size. Together with the very low p-values ($\sim 10^{-151}$ for `#shares`, $\sim 10^{-95}$ for `#tweets`), this makes us reject the null hypothesis and conclude that the differences between the two forecasting methods are statistically significant.

### 5.3 Forecasting performance on difficult videos

The difference of forecasting performance is even more noteworthy for more *difficult* videos – those that present a large exogenous shock in the forecasting period. A real example of such a video is depicted in Fig. 15(a). A video is considered to present a high exogenous shock if the exogenous stimuli series $s(t'), t' \in \text{Test}$ contains at least one point $t'$ so that $s(t') > mean(s(t^*)) + 10var(s(t^*))$, with $t^* \in \text{Train}$. 4006 videos in the Active dataset present a high exogenous shock in the testing period. MLR, even in the presence of known information about the external stimuli, largely misses the predictions (as shown in Fig. 15(b)). HIP achieves levels of forecasting performance on the high exogenous dataset similar to the entire Active dataset. Fig. 15(c) shows the distribution of absolute forecasting error for Hawkes (`#shares`) and MLR (`#shares`). Compared to Fig. 13(c), the HIP model achieves smaller errors, with a median value of 3.25%, while MLR presents an increased concentration of high errors and a median value of median value 6.5%.

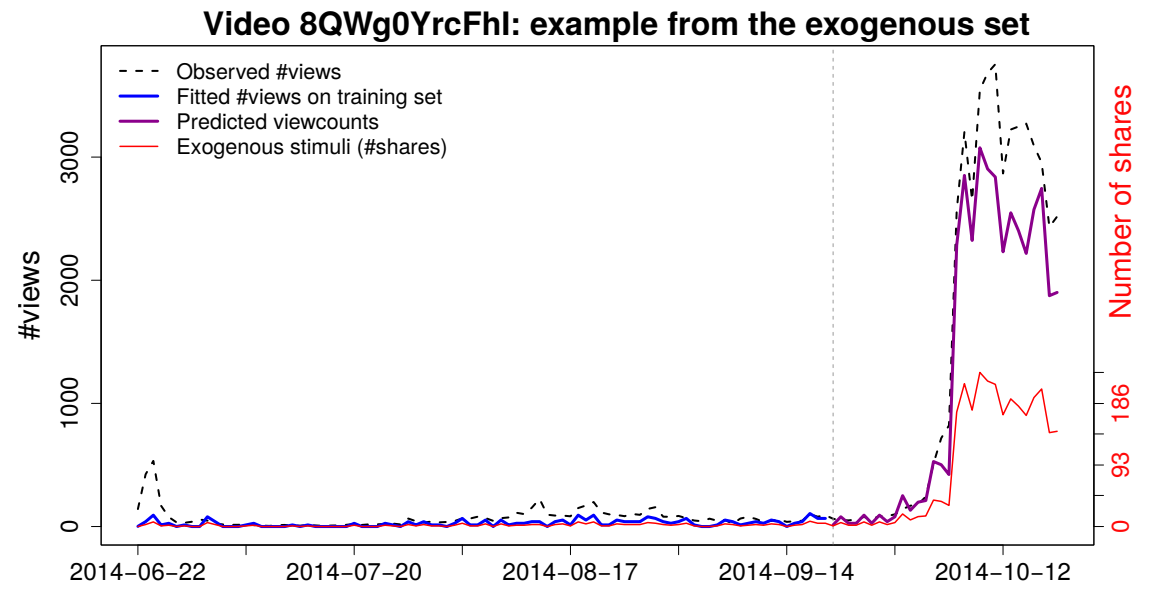

(a)

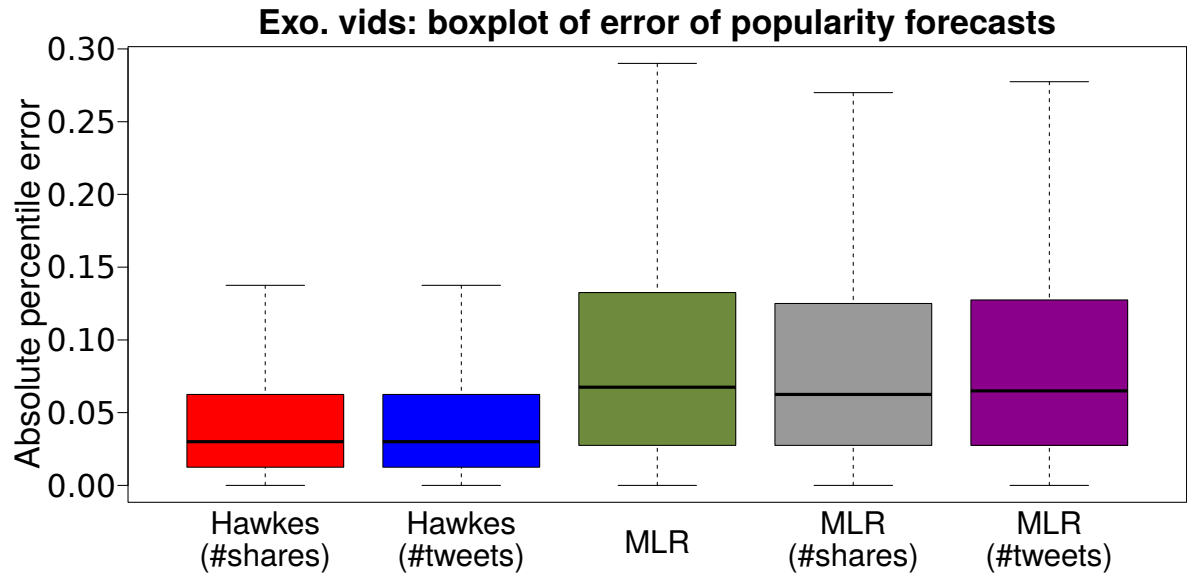

(b)

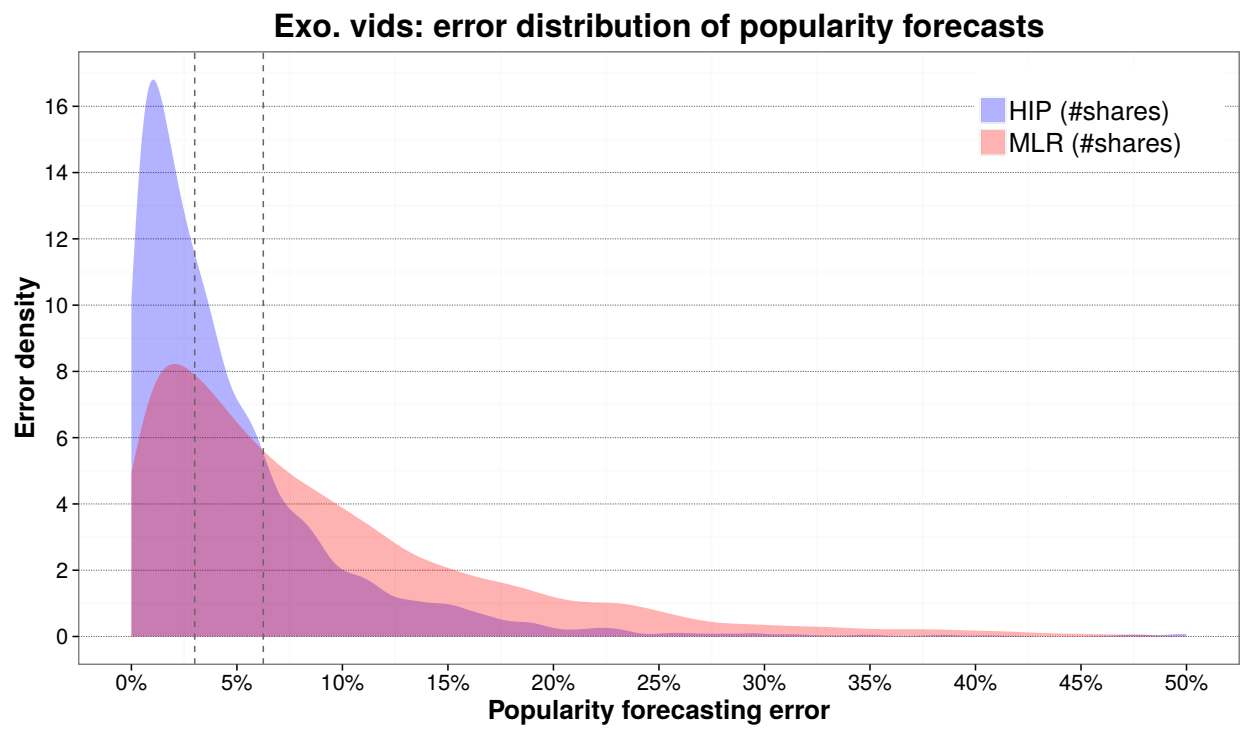

(c)

Figure 15: Forecasting performance for the Hawkes intensity model and MLR on videos presenting a high exogenous shock. (left) Depiction of a video having a high exogenous shock during the testing period. It is a relatively obscure video about Brazilian politics, which suddenly received a considerable amount of attention in October/November 2015 (more than 3 months after its upload), only to slide back into obscurity after December 2015. (right) Barplot aggregation of performances of Hawkes intensity and MLR (with different sources of external stimuli), in terms of absolute forecasting error. (bottom) Absolute forecasting error distribution for the exogenous set for Hawkes (`#shares`) (blue curve) and MLR (`#shares`) (red curve). Median values are shown with gray vertical dotted lines (3.25% for Hawkes and 6.5% for MLR).

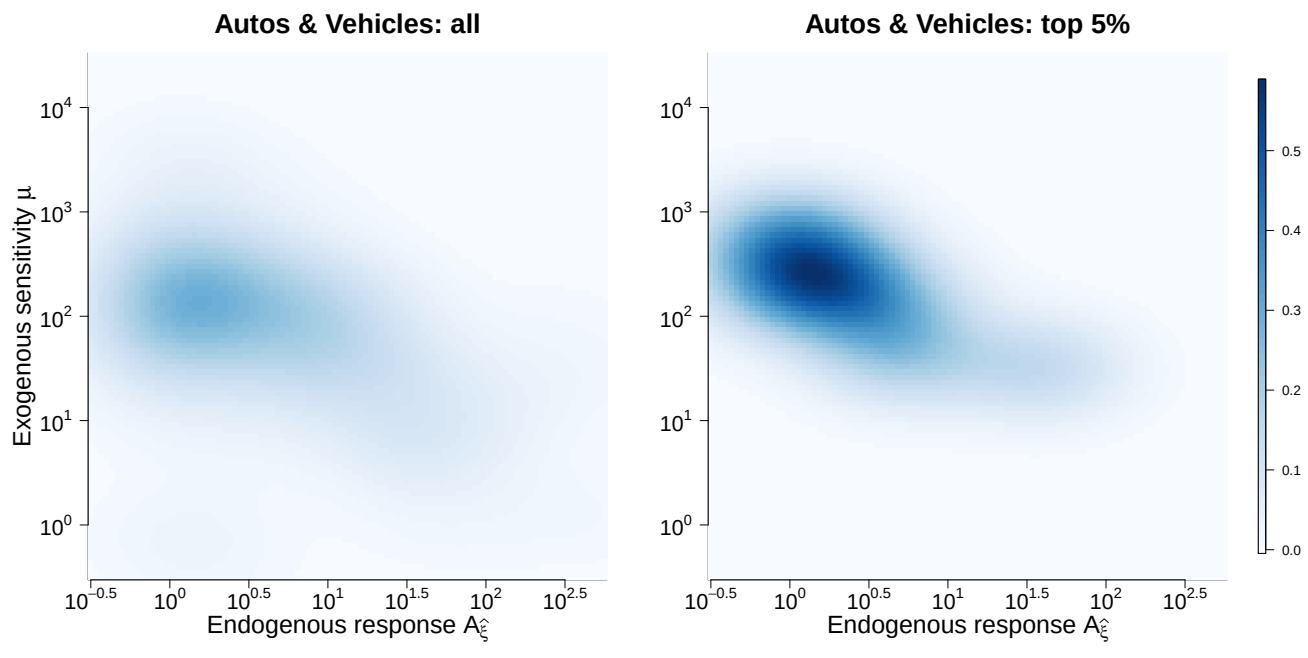

(a)

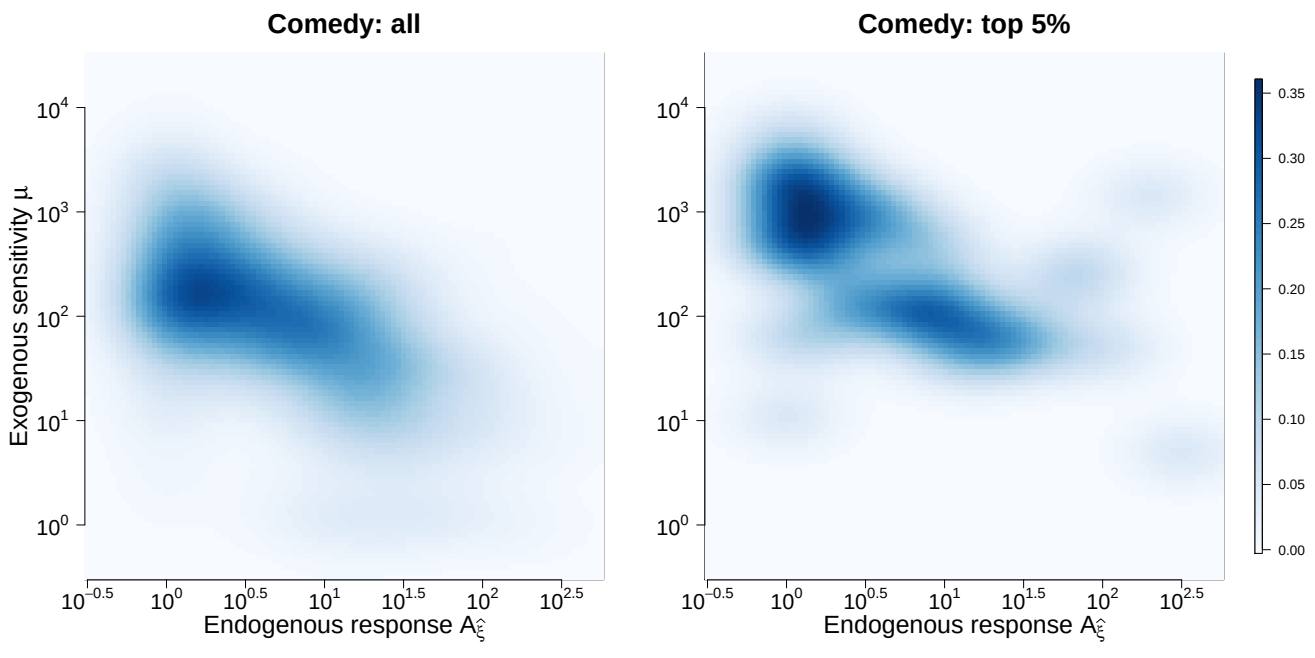

(b)

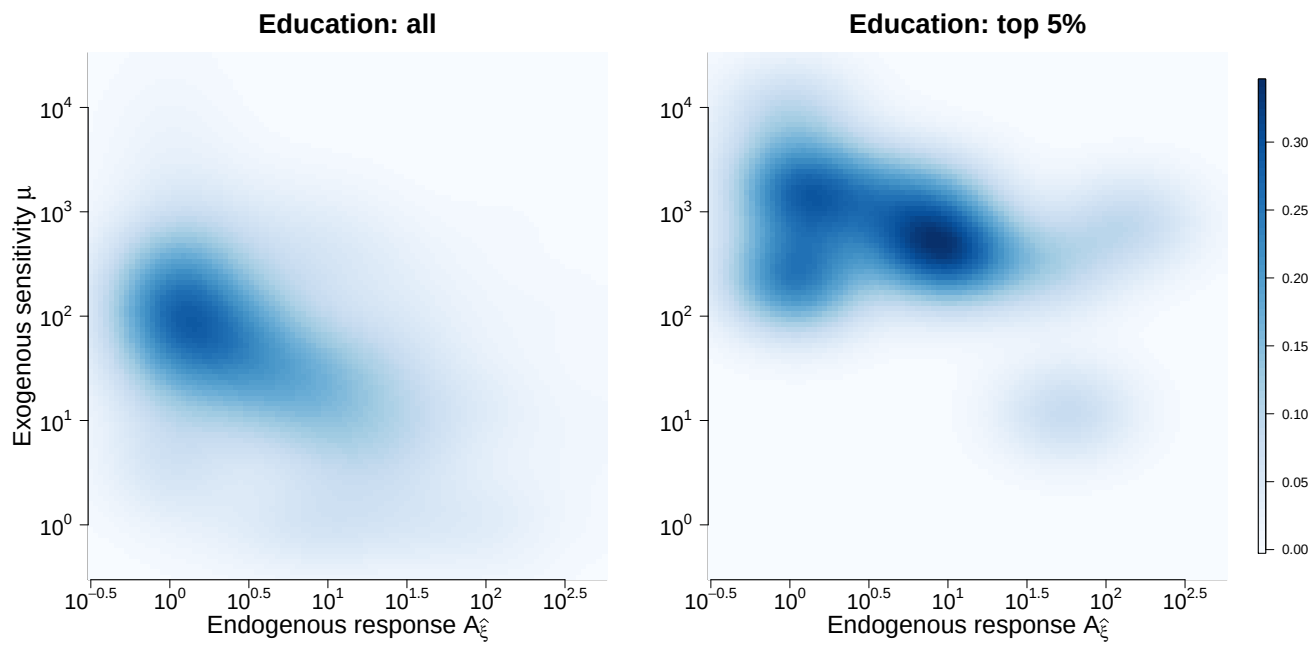

(c)

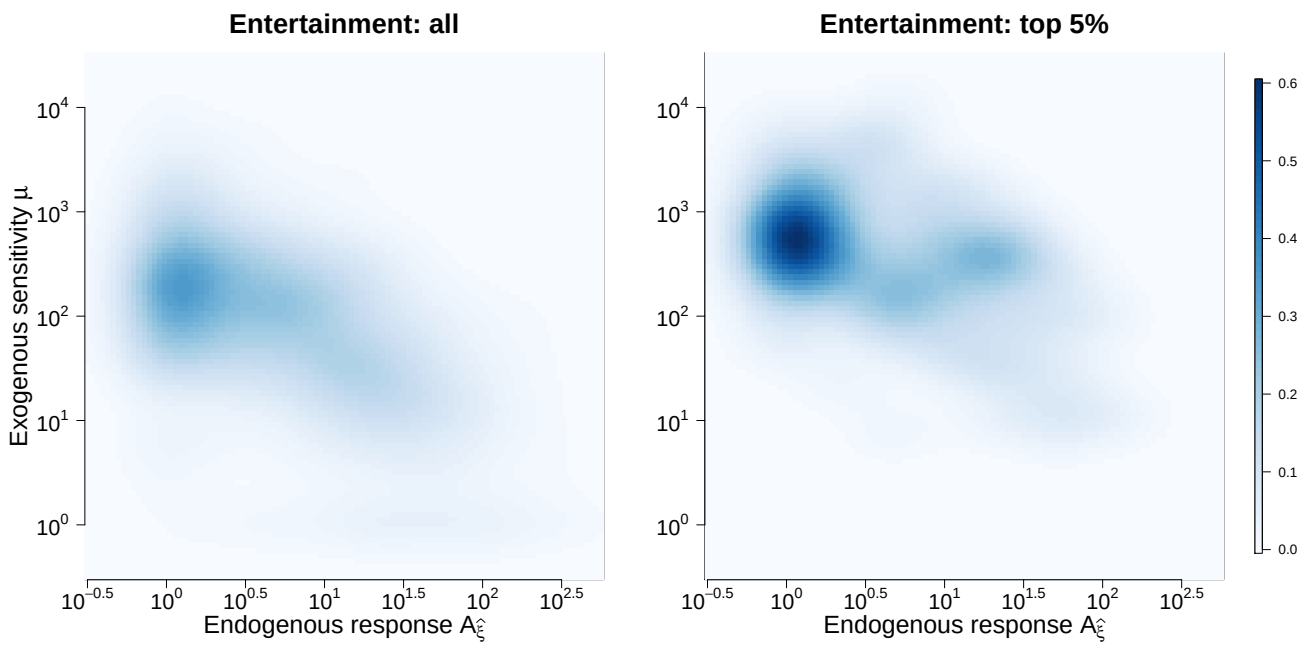

(d)

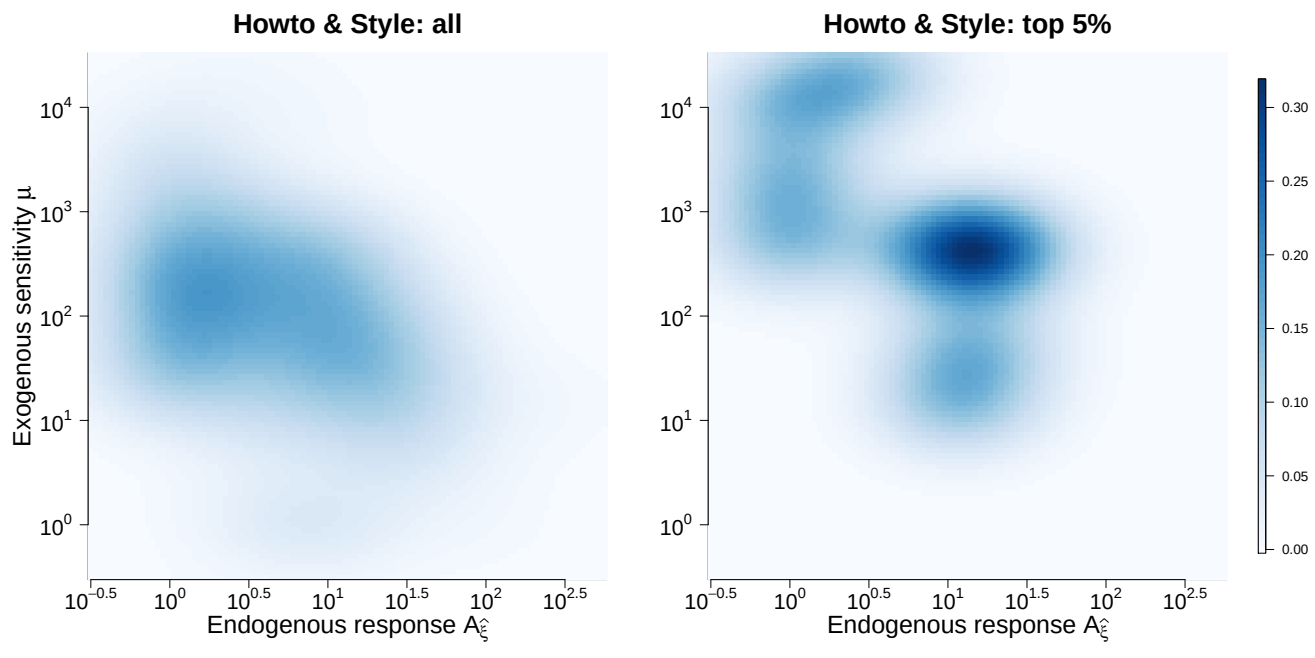

(e)

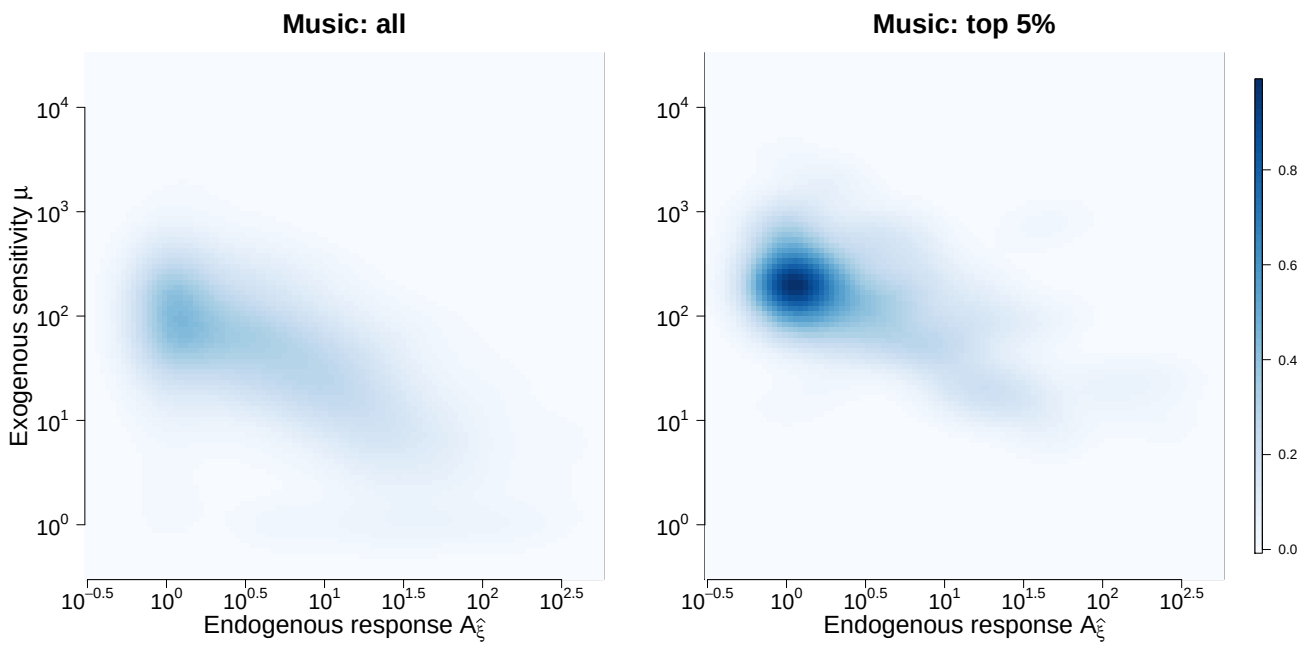

(f)

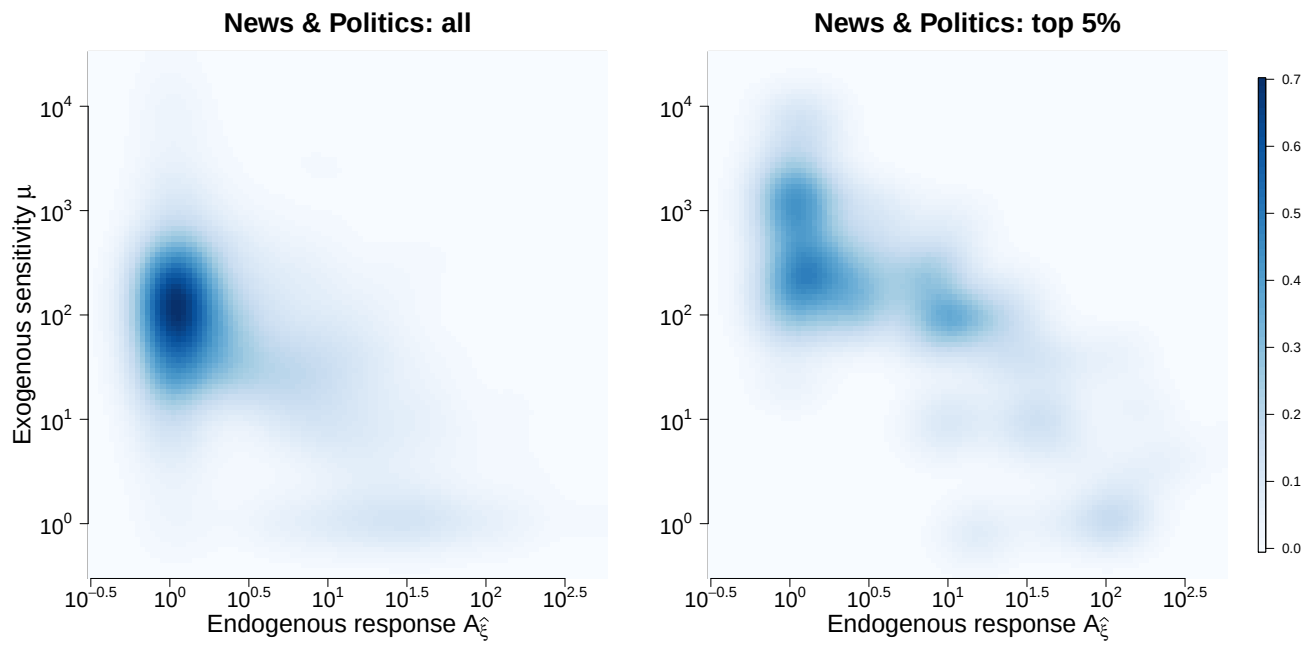

(g)

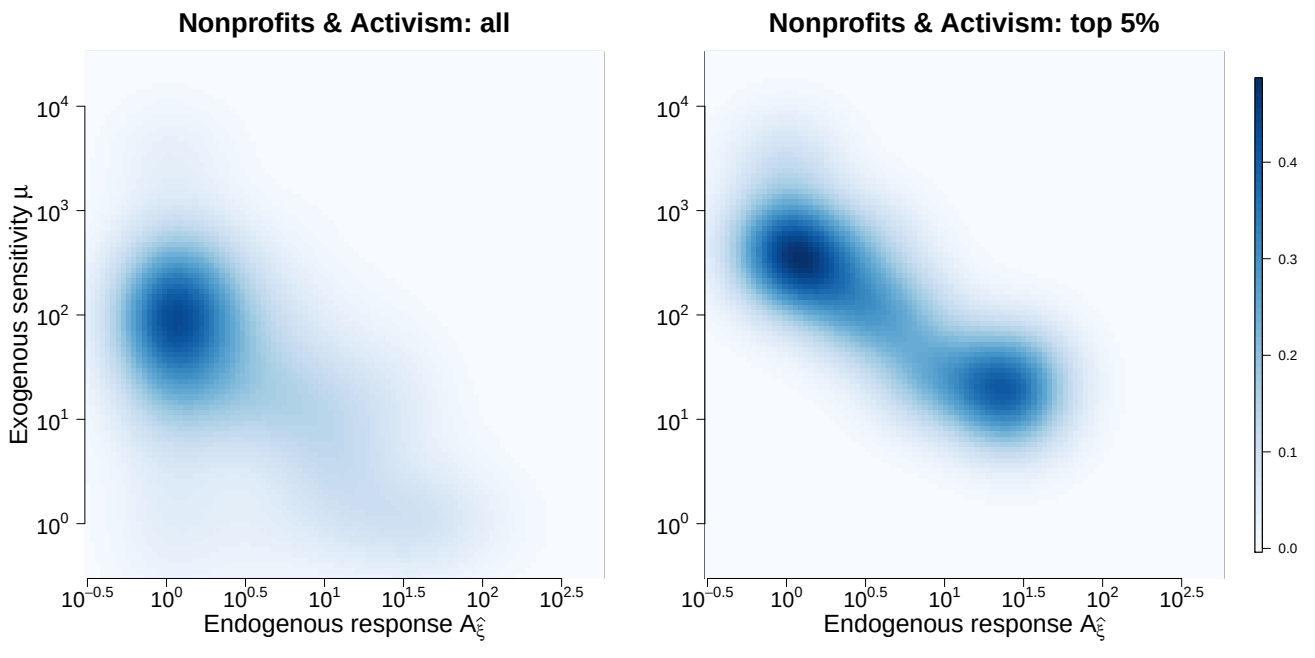

(h)

Figure 16: Pairs of 2-dimensional densities of videos on the endo-exo map, for each of the video categories in Active dataset (except for `Gaming` and `Film & Animation`, already presented in the Main Text Fig. 4). For each pair, the left heatmap represents the density distribution of all videos in the category, while the right heatmap shows the distribution of the most popular 5% videos in the category. 6 more categories are shown in Fig. 17.

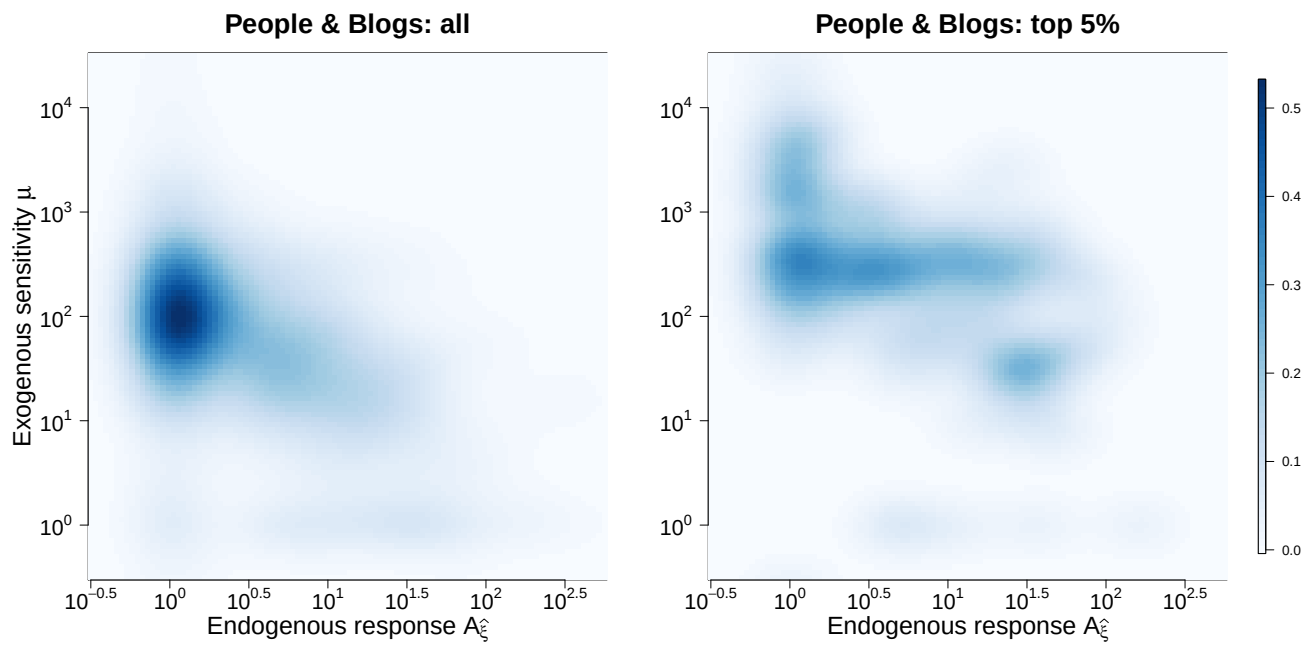

(a)

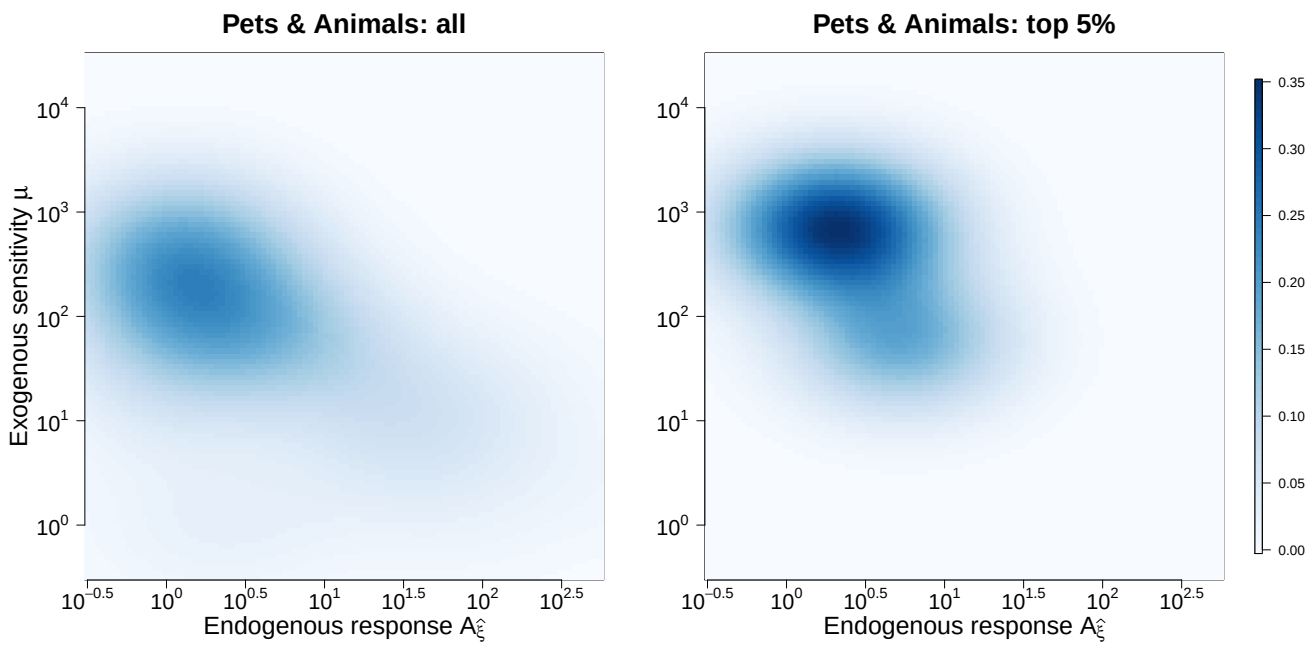

(b)

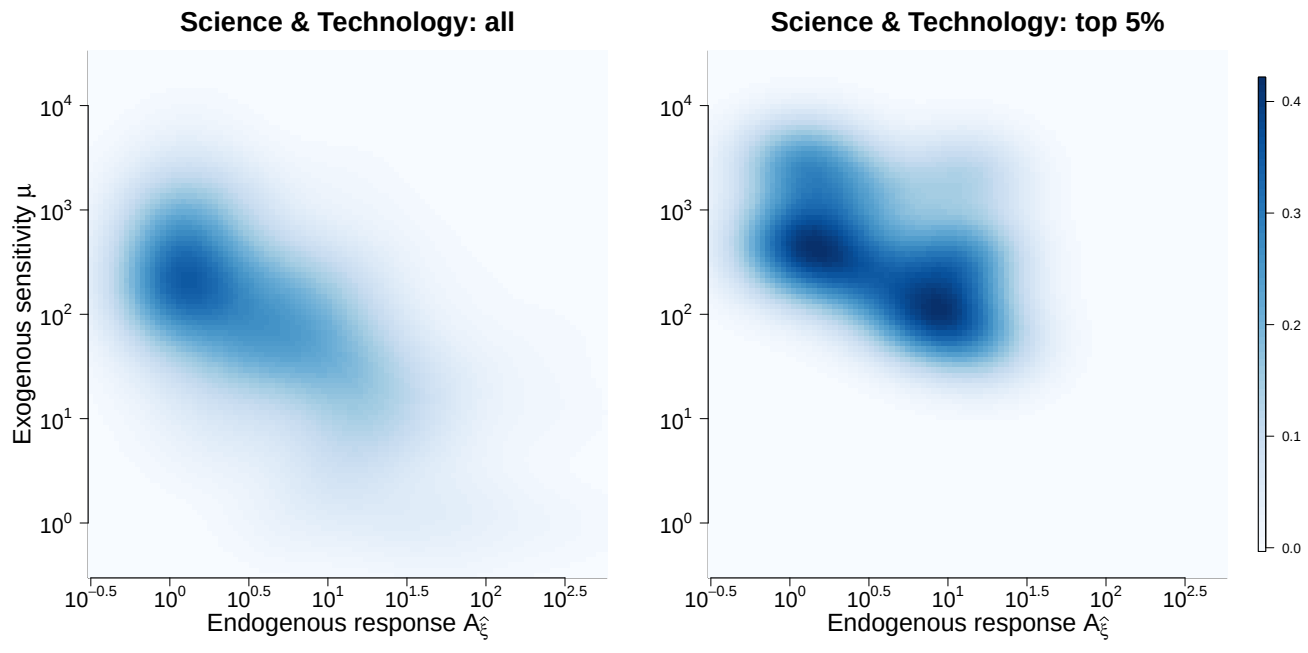

(c)

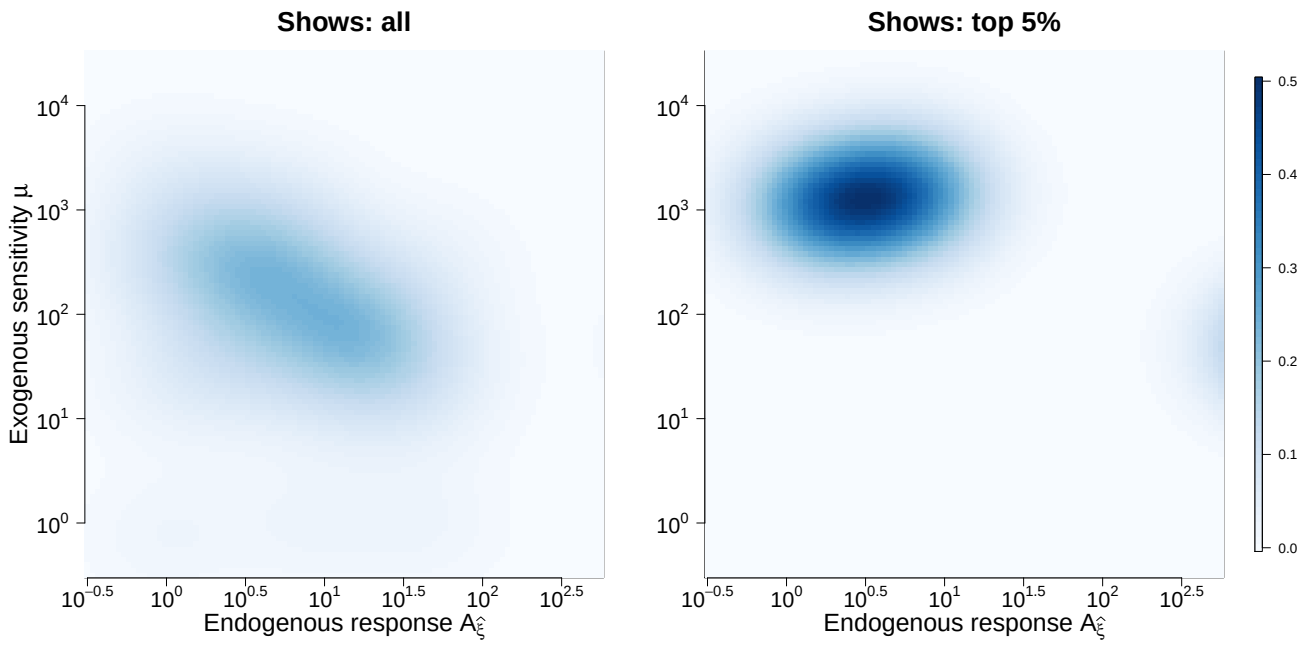

(d)

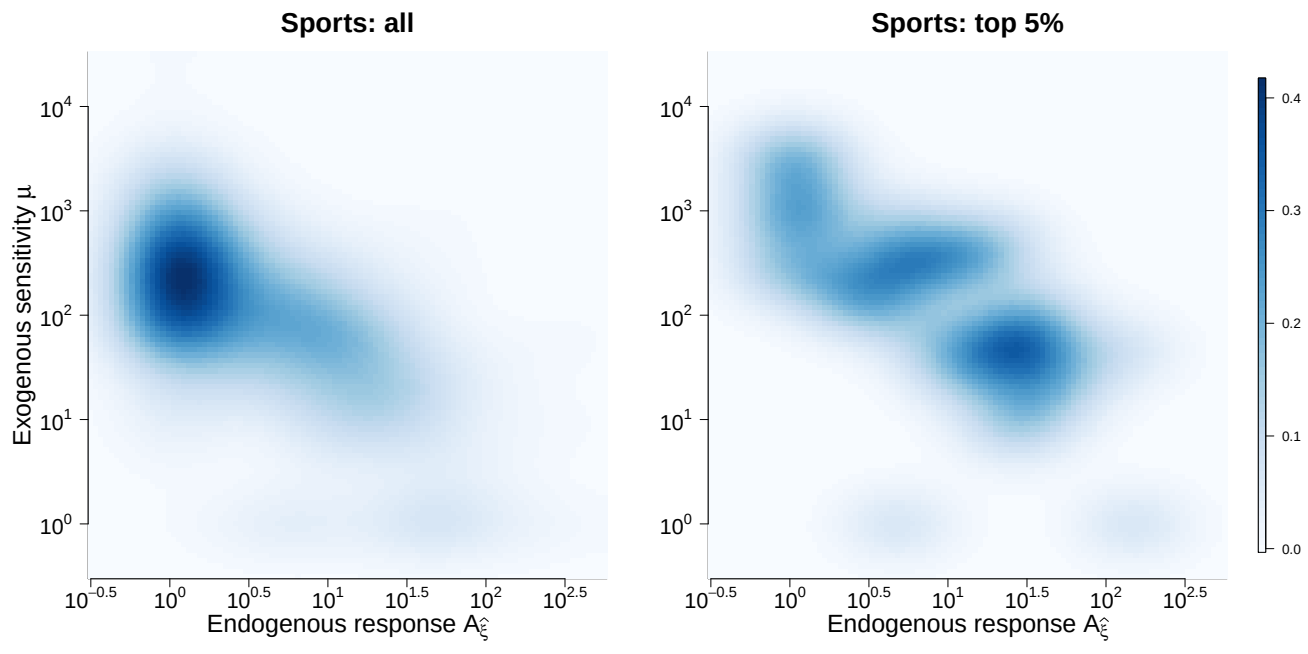

(e)

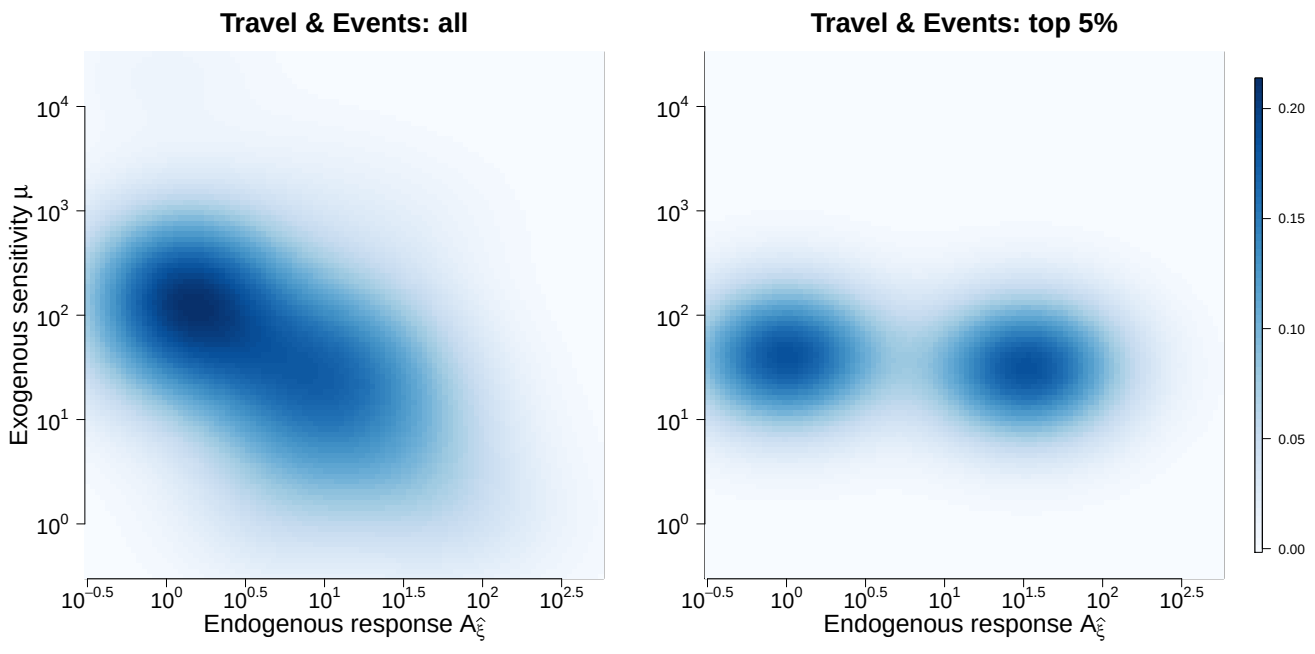

(f)

Figure 17: Fig. 16 cont'd: remaining categories.